\documentclass[structabstract]{aa}

\usepackage{txfonts}
\usepackage{graphicx}
\usepackage{natbib}
\usepackage{aalongtable}
\usepackage{gensymb}
\bibpunct[; ]{(}{)}{,}{a}{}{;}

\begin{document}

\title{Eta Carinae's 2014.6 Spectroscopic Event: \\ Clues to the Long-term
Recovery from its Great Eruption\thanks{
Based on observations made with the NASA/ESA Hubble Space Telescope, obtained [from the Data Archive] at the Space Telescope Science Institute, which is operated by the Association of Universities for Research in Astronomy, Inc., under NASA contract NAS 5-26555. These observations are associated with programmes \#7302, 8036, 8327, 8483, 8619, 9083, 9242, 9337, 9420, 9973, 11506, 11612, 12013, 12508, 12750, 13377, and 13789.
Based on observations collected at the European Southern Observatory, Chile under Prog-IDs: 60.A-9022(A), 70.D-0607(A), 71.D-0168(A), 072.D-0524(A), 074.D-0141(A), 077.D-0618(A), 380.D-0036(A), 381.D-0004(A), 282.D-5073(A,B,C,D,E), 089.D-0024(A), 592.D-0047(A,B,C). Based in part on data obtained with the
  SMARTS/CTIO 1.5m, operated by the SMARTS Consortium.} 
}

 \titlerunning{Eta Carinae's 2014.6 Spectroscopic Event} 

\author{A. Mehner\inst{1}
\and K. Davidson\inst{2} 
\and R.M. Humphreys\inst{2}
\and F.M. Walter\inst{3}
\and D. Baade\inst{4} 
\and W.J. de Wit\inst{1} 
\and J. Martin\inst{5} 
\and K. Ishibashi\inst{6}
\and T. Rivinius\inst{1}
\and C. Martayan\inst{1} 
\and M.T. Ruiz\inst{7}
\and K. Weis\inst{8}
 } 
 

\institute{ESO -- European Organisation for Astronomical Research in the Southern Hemisphere, Alonso de Cordova 3107, Vitacura, Santiago, Chile 
  \and Department of Astronomy, University of Minnesota, Minneapolis, MN 55455, USA
  \and Department of Physics and Astronomy, Stony Brook University, Stony Brook, NY 11794-3800, USA
  \and ESO -- European Organisation for Astronomical Research in the Southern Hemisphere, Karl-Schwarzschild-Stra{\ss}e 2,  85748 Garching, Germany  
  \and University of Illinois Springfield, Springfield, IL 62703, USA
  \and Division of Elementary Particle Physics and Astrophysics, Graduate School of Science, Nagoya University, Nagoya, 464-8602, Japan
  \and Universidad de Chile, Departamento Astronom\'{i}a, Casilla 36-D, Camino del Observatorio 1515, Las Condes, Santiago, Chile
    \and Astronomisches Institut, Ruhr-Universit\"{a}t Bochum, Universit\"{a}tsstrasse 150, 44780 Bochum, Germany
    }


\abstract {} 
{Every 5.5~years $\eta$~Car's light curve and spectrum change remarkably across all observed wavelength bands. These so-called spectroscopic events are likely caused by the close approach of a companion. We  compare the recent spectroscopic event in mid-2014 to the events in 2003 and 2009 and investigate long-term trends.} 
{Eta~Car was observed with \protect{{\it HST\/}} STIS, VLT UVES, and CTIO 1.5m CHIRON for a period of more than two years in 2012--2015. Archival observations with these instruments cover three orbital cycles and the events of 2003.5, 2009.1, and 2014.6. 
The STIS spectra provide high spatial resolution and include epochs during the 2014 event when observations from most ground-based observatories were not feasible.
The strategy for UVES observations allows for a multi-dimensional analysis, because each location in the reflection nebula is correlated with different stellar latitude.}
{Important spectroscopic diagnostics during $\eta$~Car's events show significant changes in 2014 compared to previous events. While the timing of the first \ion{He}{II} $\lambda$4686 flash was remarkably similar to previous events, the \ion{He}{II} equivalent widths were slightly larger and the line flux increased by a factor of $\sim$7 compared to 2003. The second \ion{He}{II} peak occurred at about the same phase as in 2009, but was stronger. The \ion{He}{I} line flux grew by a factor of $\sim$8 in 2009--2014 compared to 1998--2003. \ion{N}{II} emission lines also increased in strength. On the other hand, H$\alpha$ and \ion{Fe}{II} lines show the smallest emission strengths ever observed in $\eta$~Car. The optical continuum brightened by a factor of $\sim$4 in the last 10--15 years. \protect{ The polar spectrum shows less changes in the broad wind emission lines; the \ion{Fe}{II} emission strength decreased by a factor of $\sim$2 (compared to a factor of $\sim$4 in our direct line of sight).  The \ion{He}{II} equivalent widths at FOS4 were larger in 2009 and 2014 than during the 2003 event.}} 
{The basic character of $\eta$~Car's spectroscopic events has changed in the past 2--3 cycles; ionizing UV radiation dramatically weakened during each pre-2014 event but not in 2014. The strengthening of \ion{He}{I} and \ion{N}{II} emission and the weakening of the lower-excitation  H$\alpha$ and \ion{Fe}{II} wind features \protect{ in our direct line of sight} implies a substantial change in the physical parameters of the emitting regions. \protect{ The polar spectrum at FOS4 shows less changes in the broad wind emission lines, which may be explained by the latitude-dependent wind structure of $\eta$~Car.} The quick and strong recovery of the \ion{He}{II} emission in 2014 supports a scenario, in which the wind-wind shock may not have completely collapsed as was proposed for previous events. As the result, the companion did not accrete as much material as in previous events.  All this may be the consequence of just one elementary change, namely a strong decrease in the primary's mass-loss rate.  This would mark the beginning of a new phase, in which the spectroscopic events can be described as an occultation by the primary's wind.}


\maketitle

\section{Introduction}
\label{intro}

Giant eruptions of massive stars cause transient events that can be confounded with supernovae; this fact is one of the most important unsolved problems in stellar astrophysics. Eta~Car underwent such an eruption from 1837 to 1858 and is the only {\it supernova impostor} \citep{2005ASPC..332...47V} where the recovery from such an event can be studied in detail (see reviews and references in \citealt{2012ASSL..384.....D}). 
In the past several years $\eta$~Car's recovery process with respect to its pre-eruption brightness and the dispersal of circumstellar material has reached an epoch of rapid change. About 15 years
ago the secular brightening trend accelerated \citep{1999AJ....118.1777D, 
2004AJ....127.2352M,2006AJ....132.2717M,2010AJ....139.2056M}. Today the central star appears 6--7 times brighter in the near-ultraviolet (UV) than it did in 2000 \citep{2012ApJ...751...73M}. This requires a fundamental change in $\eta$~Car's outflow density and/or UV output if the innermost dust is being destroyed or the dust-formation rate has slowed (the most likely explanations based on available data, see \citealt{2010ApJ...717L..22M}). 

The star's brightening is accompanied by changes in the stellar wind spectrum. Optical spectra in 2009--2012 revealed a factor of 2--3 decrease in the stellar-wind emission equivalent widths since 1999 \citep{2010ApJ...717L..22M,2012ApJ...751...73M}, suggesting a decline of $\eta$~Car's mass-loss rate. The X-ray light curve is consistent with a decreasing mass-loss rate \citep{2009ApJ...701L..59K,2010ApJ...725.1528C,2010ApJ...717L..22M,2012ApJ...751...73M}.
These photometric and spectroscopic observations demonstrate that the recovery could still be ongoing. As the star returns to equilibrium, the wind structure, mass-loss rate, and spectrum evolve. This process is a unique probe of the post-eruption internal and stellar wind structure and may provide clues to the instability mechanism. On the other hand, $\eta$~Car is close to its Eddington limit and thus its intrinsic stability is precarious. The {\it amount\/} of change in $\eta$~Car's stellar wind parameters is highly uncertain, because major changes of the spectral appearance in our line of sight may be triggered by smaller changes in $\eta$~Car's stellar parameters \citep{2006AJ....132.2717M,2013MNRAS.436.3820M}.

Eta~Car's spectra and photometry show a 5.5~year cycle associated with the orbit of a companion star \citep{1997NewA....2..107D,1997NewA....2..387D}. The nature of the companion star is not well determined, but it is thought to be a 30--60~$M_\odot$ main sequence star (e.g., \citealt{2010ApJ...710..729M}). Each event shows a very rapid periastron passage of about 100~days.  This implies an orbital eccentricity above 0.8 and likely 0.85--0.90.
During periastron the separation between the two stars is only 2--3~au and this close approach of the companion star is thought to incite the so-called spectroscopic events. 
For example, the high-excitation emission lines that originate from nearby ejecta, and are probably excited by the companion, disappear for a few months. Strong \ion{He}{II} $\lambda$4686 emission appears and disappears at critical times of the X-ray light curve and may be related to a shock break-up \citep{2002ASPC..262..267D,2003ApJ...597..513S,2006ApJ...652.1563S,2006ApJ...640..474M,2011ApJ...740...80M}. The line profiles and line strengths of hydrogen and helium also show an intricate (and not well understood) behavior. Many of these phenomena occur on time scales of several months.  
Differences between the events provide valuable clues to their physics and the stellar system.
We reported the differences between the 2003 and 2009 events and their implications in \citet{2011ApJ...740...80M}. 
Periastron passages may be triggering the long-term changes observed in $\eta$~Car's light curves and spectra over the last two decades \citep{2014A&A...564A..14M}. 

The bipolar ``Homunculus'' nebula around $\eta$~Car provides us with the unique opportunity to investigate this stellar system from different directions using the reflection nebula as a giant extra-terrestrial mirror \citep{1999ASPC..179..107H,2001AJ....121.1569D}. 
\citet{2003ApJ...586..432S} showed that $\eta$~Car's wind is stronger towards the poles based on distinctive line profiles at different stellar latitudes and that the latitude-dependence of spectral features changes during events. 
Van Boekel et al.\ (2003) and \citet{2007A&A...464...87W} resolved $\eta$~Car's wind directly with interferometric observations. They found an optically thick wind with a diameter on the order of 4.3~mas ($\approx$9~au) in K-band and elongated along the direction of the Homunculus nebula. 
Recent work suggested that $\eta$~Car's wind may be more spherical than previously assumed and that the observed changes in the latitude-dependence close to periastron, but also the observed latitude-dependence at apastron, may be caused by a companion instead \citep{2010AJ....139.1534R,2012ApJ...759L...2G,2012ApJ...751...73M}. 

Our direct view of $\eta$~Car is near stellar latitude 45\degree\ \citep{2001AJ....121.1569D,2006ApJ...644.1151S}.     
A particular location in the southeast Homunculus lobe, however, provides a reflected view from almost a polar-axis direction. That location is traditionally called ``FOS4'' and was observed and discussed by \citet{2005A&A...435..303S} and \citet{2011ApJ...740...80M}.   
It is valuable for three different reasons.  (1) The spectrum of the stellar wind may depend on latitude as noted above.  (2) In most credible models, the orbit plane is approximately perpendicular to the polar axis.  Therefore major Doppler velocity variations seen at FOS4 should not represent orbital motions.   
(3)  Emitting material may be eclipsed by the primary wind during a spectroscopic event.  Near the polar axis, such eclipses very likely do not occur, or in any case must differ from our direct view.  
(One can of course imagine models with arbitrarily tilted orbits;  but only a contrived choice of parameters would cause the orbital motion to affect timing, spectra, and velocities similarly at both FOS4 and our direct view.) 
Since the FOS4 reflecting region is located about 20\,000~au from the star, it is practically equivalent to a ``view from infinity.''

With each orbit, the companion offers the unique opportunity to probe the structure of the primary's wind and circumstellar material and to assemble its evolutionary path.
We analyzed the most prominent spectral features at different stellar latitudes during the 2014 event and evaluated several competing hypotheses regarding phenomena occurring during the spectroscopic events. 
We also investigated the observed long-term trends by studying the time scales and amplitudes on which spectral changes evolve during the 2014 event compared to previous cycles. 
In Section \ref{obs} we describe the observations, followed by the results in Section \ref{results}, a discussion in Section \ref{discussion}, and the conclusions in Section \ref{conclusion}.

\section{Observations and Data Analysis}
\label{obs}

We report on new {\it Hubble Space Telescope\/} Space Telescope Imaging Spectrograph ({\it HST\/} STIS; \citealt{1998ApJ...492L..83K}) spectra obtained during the 2014 event (Table \ref{table:journal}). Some of our recent STIS observations were scheduled during the most critical epochs in 2014 Aug and Sep when observations from major ground-based facilities were not feasible. In \citet{2014arXiv1411.0695D} we have discussed a subset of these observations covering the small wavelength range around \ion{He}{II} $\lambda$4686~\AA. The STIS observations are very valuable, because they provide a homogenous data set since 1998, they include observations in the near-UV, and because of their spatial resolution, which separates the central star from the nearby ejecta. All ground-based spectra include narrow emission lines formed 0\farcs2 to 1\arcsec\ away from the star, and the amount varies because it depends on seeing, pointing, and instrument characteristics.

The STIS/CCD observations were conducted with the 52{\arcsec}$\times$0\farcs1 slit with different slit position angles in combination with the G230MB, G430M, and G750M gratings, which cover the wavelengths 2340--8080~\AA\ with spectral resolving power of $R \sim$ 5000 to 10\,000.
We used reduction methods that include several improvements 
over the normal STScI pipeline and standard CALSTIS reductions \citep{2006hstc.conf..247D}.\footnote{The reduced data from 2014 will become publicly available at http://etacar.umn.edu/ in mid-2015.} 
We extracted spectra of 0\farcs1, which at $\eta$~Car's distance of $d \approx 2.3$~kpc    
\citep{1993PASAu..10..338A,1997ARA&A..35....1D,1999ASPC..179...89M,2001AJ....121.1569D,2006ApJ...644.1151S} correspond to a projected size of about 230~au. This is much larger than the binary orbit, but small enough to exclude the Weigelt knots and similar ejecta \citep{1986A&A...163L...5W,2012ASSL..384...95H}.

We monitored $\eta$~Car regularly with the Very Large Telescope Ultraviolet and Visual Echelle Spectrograph (VLT UVES; \citealt{2000SPIE.4008..534D}) since November 2013 (Table \ref{table:journal}). In addition, we used archival data, the majority being from the $\eta$~Car campaign with UVES (P.I.\ K.\ Weis). During that campaign $\eta$~Car was regularly observed between 2002 and 2009, see \citet{2005ASPC..332..160W,2005A&A...435..303S}, and \citet{2005AJ....129.1694W} for more details on the data and results.  
The UVES spectra since 2000 provide a homogeneous data set of high spectral resolution, covering three orbital cycles. 
UVES is a high-resolution optical spectrograph, operating from 300--1100~nm. We used the DIC1 mode with central wavelengths 346+580 and the DIC2 mode with central wavelengths 437+860. Slit widths of 0\farcs4 in the blue arm and 0\farcs3 in the red arm provide spectral resolutions of 80\,000 and 110\,000, respectively. The atmospheric seeing varied from 0\farcs5--2\farcs2 with a median seeing of 1\farcs0. Observations were conducted with a slit position angle of 160\degree\ at two locations in the nebula. The first slit position is centered on the star and the second position is offset 2\farcs6 south and 2\farcs8 east of the star and is covering a location called FOS4 in the south-east (SE) nebula. These observations are an extraordinary opportunity to investigate $\eta$~Car from different directions as the star entered and exited the event in mid-2014. UVES data from the 2003 event viewed from different directions were presented in \citet{2005A&A...435..303S}.

The UVES data were reduced using the ESO UVES pipeline (version 5.3.0 and 5.4.3\footnote{The older UVES data were reduced with previous pipeline versions.}). We extracted spectra at different positions in the nebula.
The known geometry of the bipolar nebula allows us to correlate each position in the SE lobe with stellar latitude, assuming that the polar axis of $\eta$~Car is aligned with the Homunculus axis. 

As noted in Section \ref{intro}, our line of sight has a stellar latitude around 45\degree, while reflected spectra in the SE Homunculus lobe show higher latitudes.  We focus on location FOS4 near the center of the lobe, which reflects nearly a pole-on view \citep{2003ApJ...586..432S,1995AJ....109.1784D,1999A&A...344..211Z}.  The FOS4 position is 3.7\arcsec\ south and 2.5--3.5\arcsec\  east of the star.
The spectrum at FOS4 is much less contaminated by nebular emission lines from the Weigelt knots than the stellar spectrum in our direct line of sight. The reasons for this fact may be related to 
sub-arcsec structure in the circumstellar extinction \citep{1995AJ....109.1784D}.
Unlike direct observations of the star, FOS4 does {\it not\/} require high spatial resolution or precise pointing.  The form of the spectrum there varies only slightly across a 1\arcsec\ region.

The expansion of the nebula causes a velocity shift in the reflected spectra. The value depends on the location in the Homunculus lobe, see figure 4 in \citet{2001AJ....121.1569D}.  
Spectra in the nebula also show a time travel delay, because of the difference in paths between direct line of sight and the longer path from the reflection in the nebula \citep{1987A&A...181..333M,2005A&A...435..303S}. The exact time delay depends on the location in the nebula. 
We define FOS4 as the location where $\Delta \textnormal{v} = +100$~km~s$^{-1}$, 
which implies ${\Delta}t \sim 20$~days \citep{2005A&A...435..303S,2011ApJ...740...80M}. 

We obtained high resolution spectra with the CHIRON spectrograph \citep{2013PASP..125.1336T} on the CTIO 1.5m telescope with a high time sampling close to the 2014 event.
The Chiron spectrograph is a stable fiber-fed echelle designed to obtain precise radial velocities of bright objects. The fiber has a 2\farcs7 diameter on the sky. 
We used the image slicer mode, which affords a spectral resolution of $R \sim 79\,000$. 
We obtained both long (300~s, overexposing the strongest emission lines such as H$\alpha$) and short (30~s) observations of $\eta$~Car.

The Chiron images were processed at Yale prior to delivery. Echelle orders between 4600 and 8800~\AA\ were extracted and flat-fielded and a wavelength solution was provided. We calibrated the blaze function by fitting the spectrum of the spectrophotometric standard star $\mu$~Col. The flux-calibrated orders were interpolated onto a linear wavelength scale and merged. Overlapping orders were weighted by the signal-to-noise in the extracted orders.
For bright targets like $\eta$~Car the discontinuities where the orders are spliced amount to a few percent of the continuum level.

These three spectral data sets span a range of effective aperture sizes. The STIS spectra have an aperture of 0\farcs1, the UVES data set has varying apertures due to atmospheric seeing with a median of 1\arcsec, and the CHIRON spectra have an aperture of 2\farcs7 (large enough to not be much affected by the atmospheric seeing). The observed line emission can depend on the aperture size and thus the measurements need to be interpreted with some caution. Since outlying ejecta have structure at scales of 0.2--1.0\arcsec\ in observations centered on the star, we did not attempt to adjust STIS, UVES, and CHIRON measurements to the same effective aperture. 
The HST measurements have mutually consistent parameters, and likewise for CHIRON.  The UVES measurements need to be evaluated with some care because the effective aperture is slightly different each epoch.

When referring to ``phase'' in the 5.5~year cycle, we use a period of 2023.0~days. Phase = 0.0 occurs at MJD 50814 (1998 January 1), phase = 1.0 at MJD 52837.0 (2003 July 17), phase = 2.0 at MJD 54860.0 (2009 January 29), and phase = 3.0 at MJD 56883.0 (2014 August 14). We denote time within a spectroscopic event by $t$, such that $t = 0$ at MJD 54860.0, MJD 52837.0, etc. Periastron most likely occurs within the
range $t \approx -15$ to $+15$~days. This is the same timing system used by \citet{2011ApJ...740...80M,2014A&A...564A..14M}. Quoted wavelengths are air values and Doppler velocities are heliocentric.  
Velocities are not corrected for the systemic velocity of roughly $-$8~km~s$^{-1}$ (\citealt{2004MNRAS.351L..15S}; see also \citealt{1997AJ....113..335D}).

\section{Results}
\label{results}

The generally accepted view is that $\eta$~Car's spectroscopic events are caused by the close approach of a companion in an eccentric orbit. Convincing evidence comes from a combination of phenomena: 1. The primary star cannot produce enough ionizing photons to explain the high-excitation lines in the spectrum \citep{2001ApJ...553..837H,2012MNRAS.423.1623G}. 2. The first \ion{He}{II} emission peak occurred simultaneous with the X-ray decline \citep{2006ApJ...640..474M}.  3. The concept of shock breakup to explain the observed 2--10~keV X-ray behavior \citep{2002ASPC..262..267D,2003ApJ...597..513S,2006ApJ...652.1563S,2006ApJ...640..474M,2009MNRAS.394.1758P}.

During periastron the companion dives into $\eta$~Car's intrinsically variable wind, disturbing its wind and ionization structure.  This offers new insights every 5.5~years.
We present key spectral features and their evolution throughout the 2014 event and comparison to previous cycles.  
Spectral lines of $\eta$~Car's stellar wind regions can be classified into physically distinct categories. These categories have different combinations of radial velocity behavior, excitation processes, and dependences on the secondary star:                     
\begin{itemize}
   \item The low-excitation emission, such as \ion{H}{I} and \ion{Fe}{II} from the primary wind 
     \citep{2001ApJ...553..837H,2012MNRAS.423.1623G}.
   \item The higher-excitation \ion{He}{I} features. Most authors agree that these 
     are related in some way to the secondary star, but the details are 
     controversial \citep{2008AJ....135.1249H}. Part of the \ion{He}{I} emission and absorption may arise in the primary wind \citep{2001ApJ...553..837H,2012MNRAS.423.1623G}.
   \item \ion{N}{II} multiplets 2--5 in both absorption and emission.  
   They arise in the primary wind, but depend mainly 
   on UV radiation from the secondary star \citep{2011ApJ...737...70M}. 
   \item The \ion{He}{II} emission, which requires soft X-rays from the colliding-wind region \citep{2004ApJ...612L.133S,2006ApJ...640..474M,2011ApJ...740...80M,2012ApJ...746...73T}.
   \end{itemize}

\subsection{Hydrogen Balmer and \ion{Fe}{II} lines}
\label{sec:halpha}

\begin{figure}
\centering 
\resizebox{\hsize}{!}{\includegraphics{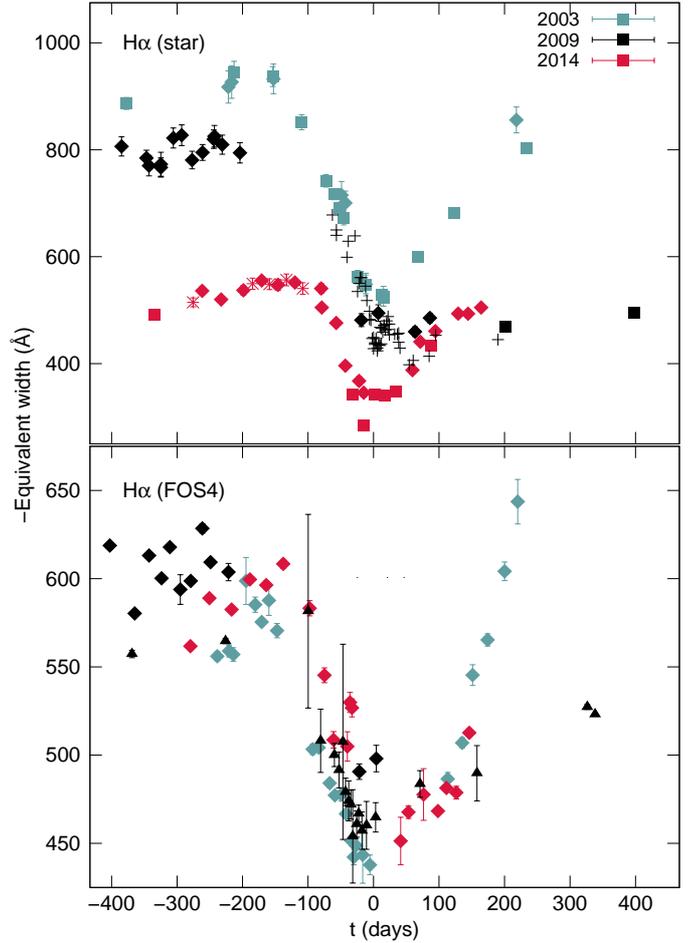}}
    \caption{Equivalent width of the H$\alpha$ emission during the 1998, 2003, 2009, and 2014 events (squares: {\it HST\/} STIS, diamonds: VLT UVES, triangles: Gemini GMOS, asterisks: CTIO SMARTS 1.5m CHIRON, plusses: \citealt{2010AJ....139.1534R}). The timing of the equivalent widths at FOS4 are corrected for the additional light travel time ($t_{\textnormal\scriptsize{FOS4}} = t+18$~d).}
     \label{fig:halphastarfos4}
\end{figure}

The equivalent width of H$\alpha$ in our line of sight has decreased progressively over the last two decades. Despite this, the minimum values during the events of 2003 and 2009 were similar. In 2014, however, the equivalent width reached its lowest value ever observed in this star (Figure \ref{fig:halphastarfos4}, upper panel). While H$\alpha$ decreased in equivalent width, the apparent line flux was a factor of $\sim$2 larger in the period 2009--2014 compared to 1998--2003. This is because the apparent continuum has brightened, partly because of decreasing circumstellar extinction.  Since the extinction factor is poorly known, equivalent width is a better-defined measure of the line's intrinsic behavior.\footnote{The STIS spectra are flux-calibrated. However, STIS observations of H$\alpha$ require very short integration times and shutter effects lead to flux discrepancies on the order of a few percent. Our ground-based observations are not flux-calibrated and thus equivalent width is the only available measure for the line strengths.} This statement applies to other spectral features as well.

It is important to note that H$\alpha$ and some of the \ion{Fe}{II} lines originate in a much larger region than anything else discussed here. Their emission zones extend beyond  $r \sim 30$~au \citep{2001ApJ...553..837H}, compared to the binary separation $r \lesssim 3$~au at periastron. In most models one can visualize the low-excitation emission region as a very diffuse prolate spheroid, with a conspicuous equatorial ``cavity'' on one side due to the secondary star's fast wind. Note also that H$\alpha$ is remarkably insensitive to gas density and temperature, unlike \ion{Fe}{II} and most other emission lines.

At FOS4, the H$\alpha$ equivalent width has been mostly constant over the last decades (outside the events). The decline during the 2014 event matched the observations in 2003 and 2009.
In the past, the difference in equivalent widths in direct view of the star and at FOS4 was puzzling and was interpreted as higher dust extinction in our line of sight \citep{1992A&A...262..153H}. Now, the equivalent width at FOS4 is larger than in our line of sight. Very likely this effect is related to latitudinal viewing angle -- i.e., FOS4 reflects the stellar wind spectrum as seen from a near-polar direction. This makes it special in a bipolar wind model \citep{2003ApJ...586..432S}, a wind-cavity model induced by the secondary star \citep{2012ApJ...759L...2G}, or any synthesis of these concepts.

\begin{figure*}
\centering 
\resizebox{\hsize}{!}{\includegraphics{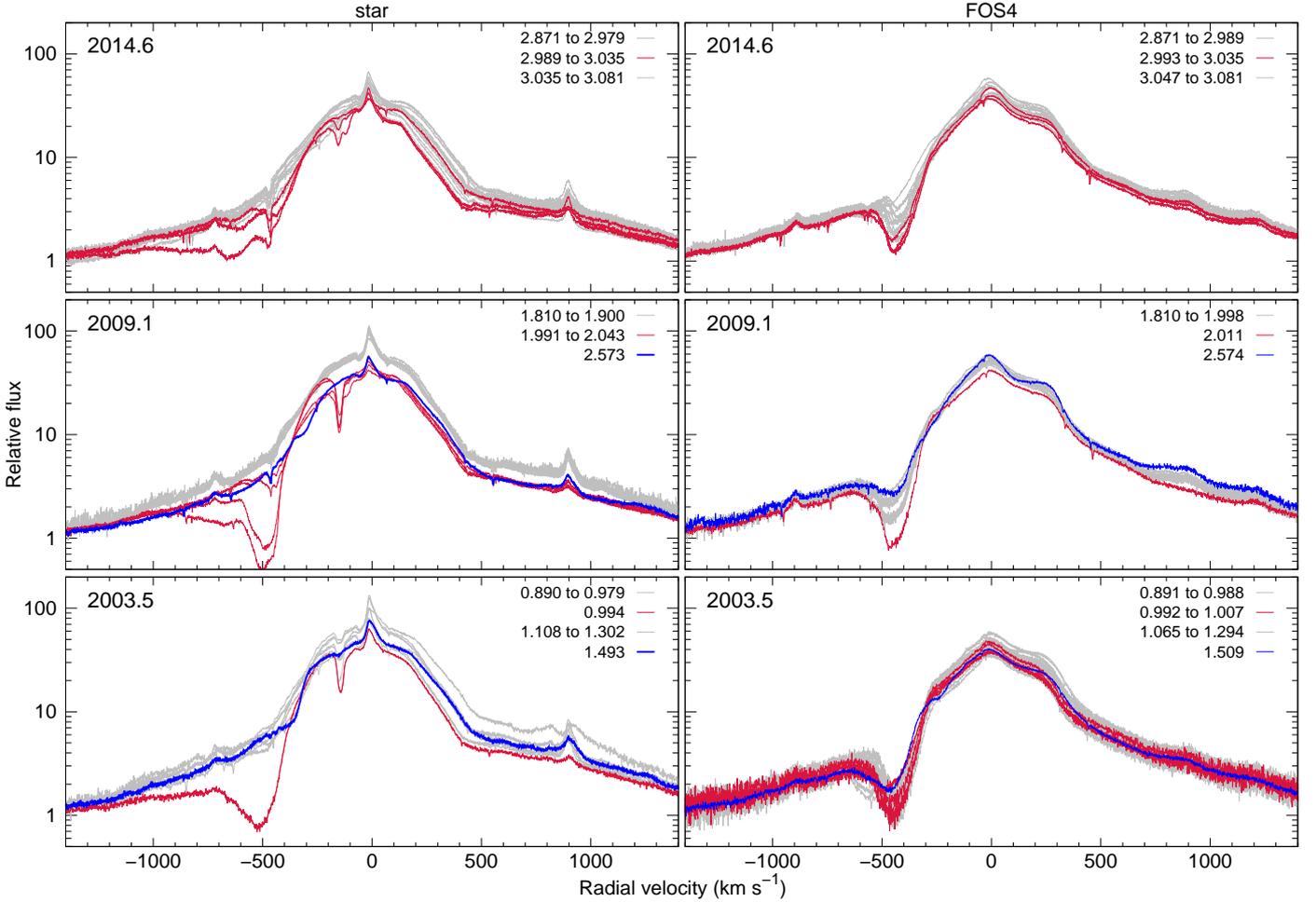}}
    \caption{Line profile changes in H$\alpha$ during the 2003, 2009, and 2014 events observed with UVES in direct view to the star and at FOS4. The spectra at FOS4 are corrected for the velocity shift of $\Delta \textnormal{v} = 100$~km~s$^{-1}$, see Section \ref{obs}. The key indicates the corresponding phases of the displayed spectra, see last paragraph in Section \ref{obs}. Red spectra indicate when notable changes in the P~Cyg absorption occurred. The blue tracings show the mid-cycle spectra when the companion star is at apastron and its influence on $\eta$~Car's wind structure is minimal. The spike atop the H$\alpha$ profile comes from older low-velocity ejecta and not from the present-day stellar wind.}
     \label{fig:halphaprofile}
\end{figure*}

For most of $\eta$~Car's cycle there is no H$\alpha$ P~Cyg profile in the spectrum of the star in our line of sight, while the spectra at FOS4 show P~Cyg absorption (Figure \ref{fig:halphaprofile}). This was seen as evidence for a latitude-dependent primary wind \citep{2003ApJ...586..432S}. \citet{2012ApJ...759L...2G} suggested that the observed latitudinal dependence of H$\alpha$ line profiles are dominated by the companion star and the associated wind cavity. Figure \ref{fig:halphaprofile} shows how the H$\alpha$ profile changes around the events in 2003, 2009, and 2014. In our line of sight, P~Cyg absorption develops at a phase of $\sim$0.900 and is present until a phase of $\sim$1.100. The P~Cyg absorption profile was more structured in 2009 and 2014 than in 2003. The P Cyg absorption velocity at FOS4, roughly $-450$~km~s$^{-1}$ relative to the emission peak, is interesting because it is not much different from our direct view of the star. In simple polar-wind scenarios, one normally expects outflow velocities to depend strongly on latitude \citep{2003ApJ...586..432S,1996ApJ...472L.115O}.
Absorption extends to $-1000$~km~s$^{-1}$ at FOS4, but a deep minimum occurred near $-500$~km~s$^{-1}$ before 2014.  In 2014 it had largely vanished, being replaced by a more complex and shallower structure.

At FOS4, H$\alpha$ P~Cyg absorption is present throughout the cycle, but has been weakening relative to the continuum each cycle. During the events, the absorption component strengthens, but the line profile changes are much less pronounced than in our direct line of sight (Figure \ref{fig:halphaprofile}; see also \citealt{2005AJ....129.1694W}).

A narrow absorption feature near $-144$~km~s$^{-1}$ in the Halpha profile regularly appears close to periastron passages and was first noted during the 1981.5 event \citep{1982A&A...111..375M}. 
It disappears in the year(s) afterwards \citep{1984ApJ...285L..19R,1998A&AS..133..299D,1999ASPC..179..227D,2005AJ....129..900D,2010AJ....139.2056M,2010AJ....139.1534R}. 
It reappeared in our data between 2014 July 22 and 29. 
This absorption feature may indicate unusual nebular physics far outside the wind \citep{2005A&A...435..183J}. \citet{2010AJ....139.1534R} proposed that this anomalous absorption originates in the nearby wall of the Little Homunculus \citep{2003AJ....125.3222I} and that a change in ionizing radiation by the secondary star is the cause for its sudden appearance shortly before periastron passages.

\begin{figure}
\centering 
\resizebox{\hsize}{!}{\includegraphics{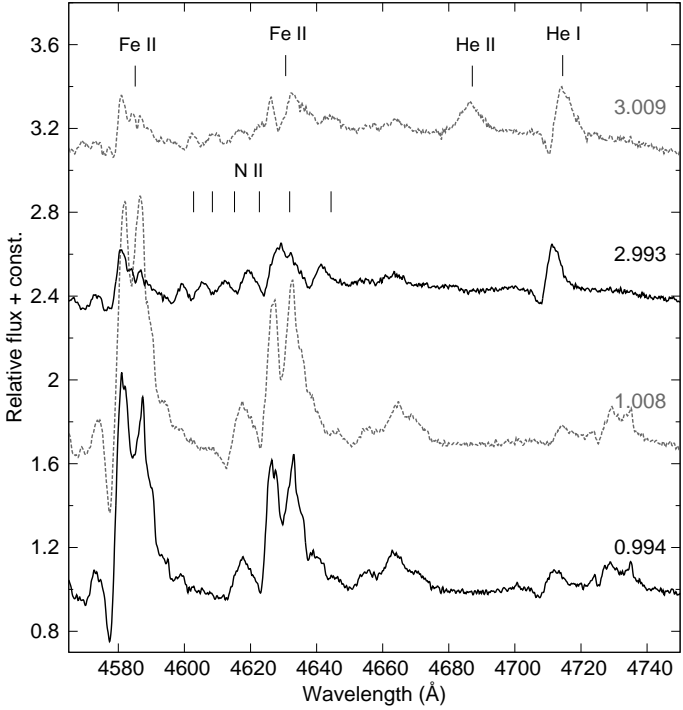}}
   \caption{STIS spectral tracings of the central star (in our direct line of sight) including the \ion{Fe}{II} $\lambda$4585,4631, \ion{He}{I} $\lambda$4714, and \ion{N}{II} $\lambda$4601--4643 lines at similar phases during the 2003 and 2014 events. The \ion{Fe}{II} lines have almost disappeared in 2014. The  \ion{He}{I} and \ion{N}{II} lines have increased in emission strengths. They show similar velocity variations with phase. \ion{He}{II} is absent in 2003 at phase 1.008, but strongly present in 2014 at phase 3.009.}
     \label{fig:compare4706}
\end{figure}

\begin{figure}
\centering 
\resizebox{\hsize}{!}{\includegraphics{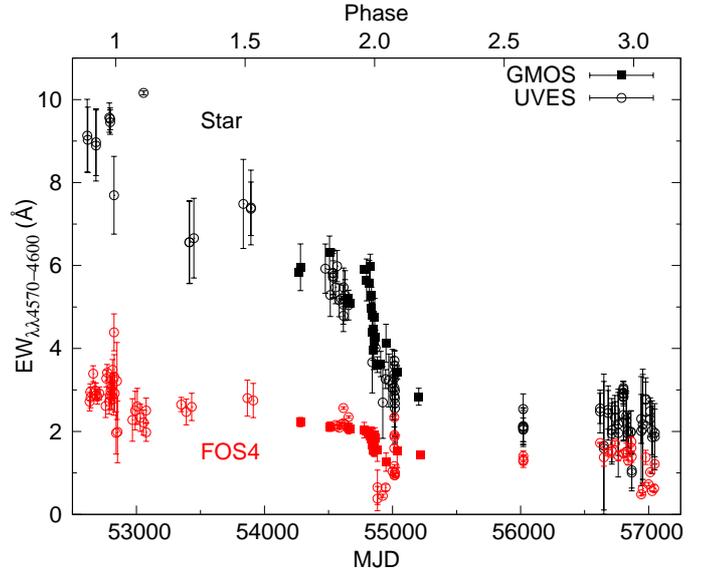}}
   \caption{Equivalent width of the broad \ion{Fe}{II} blend at 4570--4600~\AA\ in our direct line of sight (black symbols) and at FOS4 (red symbols) with Gemini GMOS and VLT UVES in 2002--2015. This is an updated version of figure 6 in \citet{2012ApJ...751...73M}. The emission in our direct view of the star decreased by a factor of $\sim$4, at FOS4 by only a factor of about $\sim$2 between 2003 and 2015.}
     \label{fig:FeII4585_EW}
\end{figure}

The \ion{Fe}{II} emission lines show decreasing equivalent widths both in direct view and at FOS4 over the last cycles (see \citealt{2012ApJ...751...73M} and Figures \ref{fig:compare4706}, \ref{fig:FeII4585_EW}, and \ref{fig:hei4714_tracing}).  The broad \ion{Fe}{II} emission at 4570--4600~\AA\ has weakened since 2003 by a factor of $\approx 4$ in our direct line of sight and by a factor of $\approx 2$ at FOS4.  The emission almost disappeared during the 2014 event.  Even though the 15-year \ion{Fe}{II} decline is relatively weaker at FOS4, it would be sufficient to indicate a trend in the stellar wind properties even if we had no direct-view spectra.

\subsection{\ion{He}{I} lines}

\begin{figure}
\centering 
\resizebox{\hsize}{!}{\includegraphics{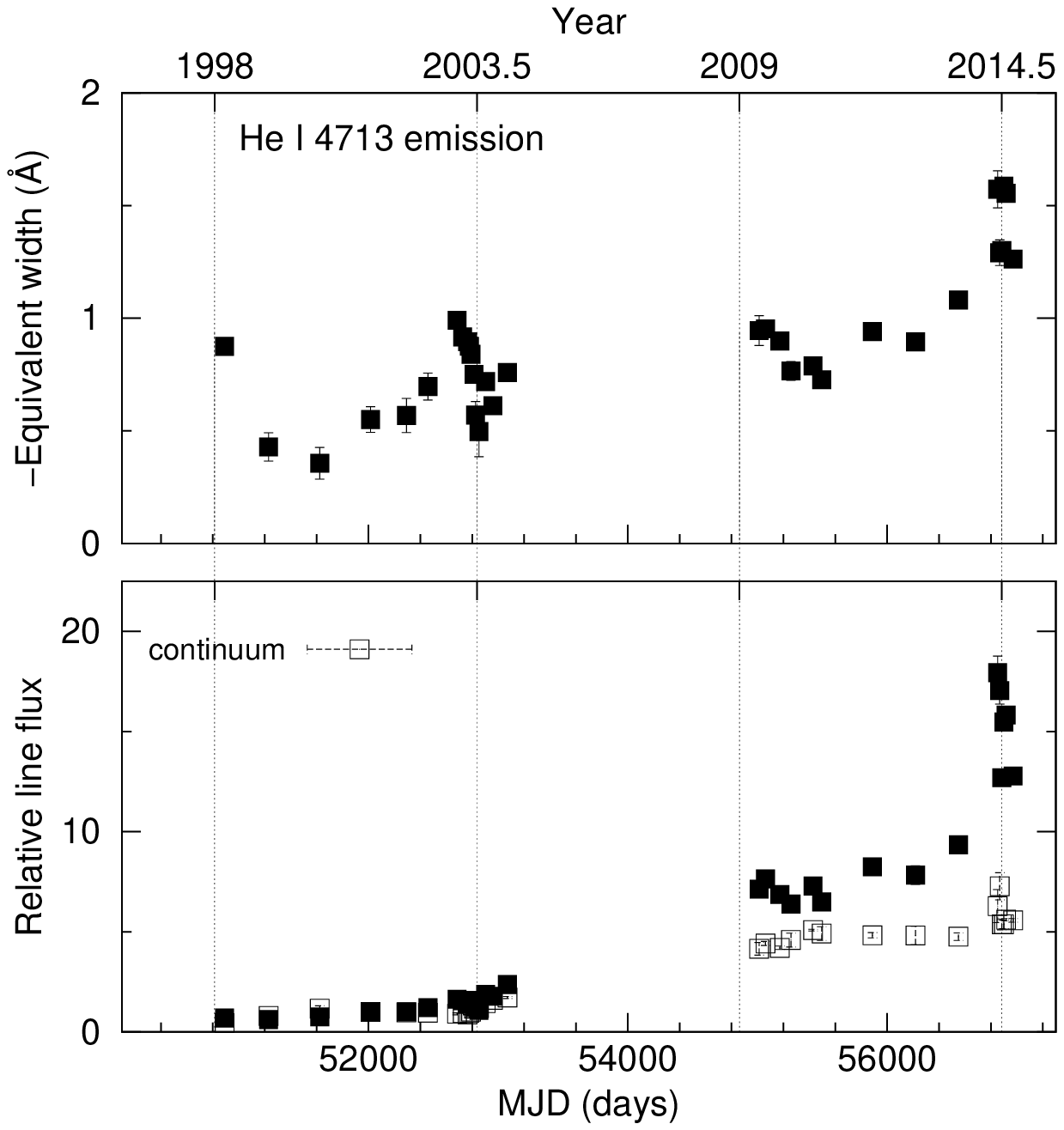}}
   \caption{Equivalent width and relative line flux of the \ion{He}{I} $\lambda$4713 emission in our direct line of sight in STIS data since 1998 March (filled squares). The line was integrated between $\sim$4711.5~\AA\ and 4721.5~\AA\ (we varied the blue side of this range somewhat to exclude the absorption component) with a continuum interpolated between 4601--4611~\AA\ and 4739--4742~\AA. The open squares show the continuum variation between $\sim$4711.5~\AA\ and 4721.5~\AA.  Phases 0.0, 1.0, 2.0, and 3.0 are indicated with vertical lines. The relative line flux and continuum were normalized to unity at phase 0.6. The \ion{He}{I} $\lambda$4713 line flux increased by more than a factor of $\sim$8 in 2009--2014 compared to 1998--2003, while the continuum increased by a factor of $\sim$5. During the 2014 event, the line became tremendously bright.}
     \label{fig:hei4714}
\end{figure}

\begin{figure}
\centering 
\resizebox{\hsize}{!}{\includegraphics{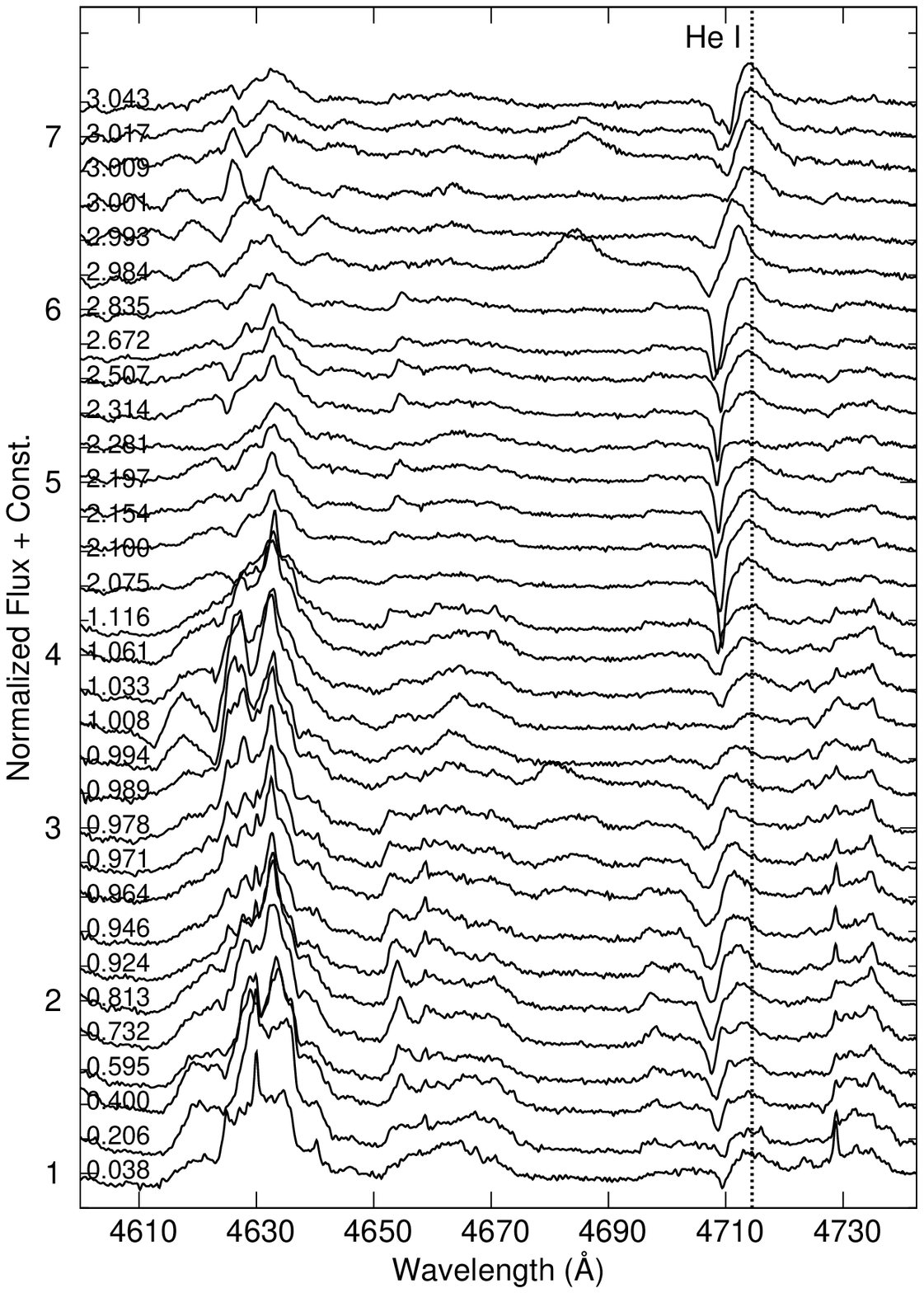}}
   \caption{ STIS spectral tracings of the central star showing the \ion{He}{I} $\lambda$4714 emission and absorption in our direct line of sight from 1998--2014. Phase in the 5.5~yr cycle are indicated next to each tracing, see Section \ref{obs}. \ion{He}{I} displays a modulation with the orbital cycle regarding the velocities and emission and absorption strengths. For example, during each event the absorption component vanishes. A long-term trend is also evident: the \ion{He}{I} absorption strength increased relative to the continuum since 1998. The Figure also shows the decrease of the \ion{Fe}{II} $\lambda$4631 emission and the appearance and disappearance of the \ion{He}{II} $\lambda$4686 emission.}
     \label{fig:hei4714_tracing}
\end{figure}

The \ion{He}{I} lines depend on photoionization by the companion star by photons with energies $> 24.6$~eV. \ion{He}{I} emission and absorption exhibit similar velocity behavior as the \ion{He}{II} emission \citep{2007ApJ...660..669N,2011ApJ...740...80M}. 
The velocity variations observed at FOS4 suggest that the helium lines are not simply related to orbital motion of the secondary star, if we assume that the orbit inclination is $i \sim$ 40--45$^{\circ}$ like the Homunculus mid-plane.  
The observed velocity changes may represent flows of highly ionized material, modulated by the secondary star. 
The details are model-dependent: e.g., \citet{2007ApJ...660..669N,2008MNRAS.386.2330D,2008AJ....135.1249H,2007NewA...12..590K,2008MNRAS.390.1751K,2006ApJ...640..474M,2001ASPC..233..173D}.  

The \ion{He}{I} emission lines dramatically increased in strength over the last 3 cycles (Figures \ref{fig:compare4706} and \ref{fig:hei4714}). Figure \ref{fig:hei4714} shows the equivalent width and the line flux of the \ion{He}{I} $\lambda$4713~\AA\ emission in STIS spectra. The equivalent width in 2009--2014 was by a factor of about 2 larger than the equivalent width in 1998--2003. The line flux showed a tremendous increase of more than a factor of $\sim$8 in 2009--2014 compared to 1998--2003 and the line became very bright during the 2014 event. This fact is especially remarkable because $\eta$~Car's spectroscopic events were originally defined by a {\it weakening\/} of high-excitation emission  \citep{1984A&A...137...79Z}.

The \ion{He}{I} P~Cyg absorption strength relative to the continuum has been increasing since 1998 (Figure \ref{fig:hei4714_tracing}). In addition, an orbital modulation of the absorption strength is observed. The strongest absorption occurs several months before and after the events, while it disappears completely during the events. The absorption became strongest (relative to the continuum) in our observations at phase 2.835. As expected from previous events, the absorption then weakened closer to the 2014 event and had disappeared at phase 3.001. The \ion{He}{I} absorption reappeared in our data at phase 3.009 and has been increasing again in strength since. Based on the orbital modulation observed for the previous cycle, we expect the absorption to increase until a phase of $\sim$3.100, before decreasing again.
To some extent, the cyclic \ion{He}{I} absorption behavior can be qualitatively understood in terms of the expected ionization-zone shapes.  See  Figure 5 in \citet{2012ApJ...751...73M}, and the related discussion there.

\subsection{\ion{N}{II} emission}

In \citet{2011ApJ...737...70M} we reported on the \ion{N}{II} multiplet at $\lambda\lambda$5667--5711~\AA. 
The \ion{N}{II} lines exhibit radial velocity variations similar to the helium lines (\citealt{2011ApJ...737...70M}; see also Figure \ref{fig:compare4706}).  The \ion{N}{II} lines and their velocity shifts are also seen in the reflected polar spectrum at FOS4.
The lines likely depend on a form of excitation by the hot secondary star, but in different regions than the \ion{He}{I} lines.
In contrast to \ion{He}{I}, the \ion{N}{II} lines depend on photoexcitation at energies of $\sim$18.5~eV.
The \ion{N}{II} features should arise primarily in regions of the primary wind that are close to the secondary star, and, therefore, close to the He$^+$ zones and the apex of the wind-wind collision zone.  

During the 2014 event, the broad emission and absorption lines of the \ion{N}{II} multiplet at $\lambda\lambda$4601--4643~\AA\ became very strong and dominated the spectrum around $\lambda$4600~\AA\ in STIS data at phase 2.993 (Figure \ref{fig:compare4706}). This was facilitated by the extreme weakening of the \ion{Fe}{II} $\lambda\lambda$4584,4629 lines. Like the bright \ion{He}{I} lines, this development is almost opposite to the spectroscopic events seen decades ago.  On those occasions, the EUV radiation temporarily declined or even disappeared  \citep{1984A&A...137...79Z}. The two \ion{N}{II} multiplets mentioned here require strong EUV radiation.

\subsection{\ion{He}{II} $\lambda$4686 emission}
\label{sec:heII}

The \ion{He}{II} $\lambda 4686$ emission line appears only briefly at certain stages during the events  \citep{2004ApJ...612L.133S,2006ApJ...640..474M,2011ApJ...740...80M,2012ApJ...746...73T}. 
This feature provides important diagnostics to the nature of $\eta$~Car's periastron passages, because of its possible connection to a break-up of the wind-wind shock.   
The $\lambda$4686 emission appears and increases in strength, when the 2--10~keV X-rays decline (\citealt{2006ApJ...640..474M}; Figure \ref{fig:heIIxray}). When the X-ray emission enters the minimum, the \ion{He}{II} emission disappears. A second, weaker occurrence of \ion{He}{II} emission is observed shortly afterwards, maybe related to the reappearance of the X-rays \citep{2011ApJ...740...80M}. 

The \ion{He}{II} $\lambda 4686$ line is by far the highest ionization feature in $\eta$~Car's UV to infrared (IR) wind spectrum. The transition is probably populated via photoionization and subsequent He$^{++}$ $\rightarrow$ He$^+$ recombination. It requires a temporary source of He$^+$-ionizing photons (with energies $> 54$~eV).
The most plausible energy source are soft (54--500~eV) X-rays, produced in the shocked gas of the primary wind. The \ion{He}{II}~$\lambda 4686$ emission likely originates in $\eta$~Car's wind close to the wind-wind collision (WWC) and/or in the cold, dense post-shock $\eta$~Car wind \citep{2006ApJ...640..474M,2007MNRAS.378..309A,2011ApJ...740...80M,2012ApJ...746...73T}. 
The supply of soft X-rays can temporarily rise to very high levels if the fast secondary-wind shock becomes unstable like the primary-wind side. In that case the entire wind-wind interface can disintegrate and collapse, and a chaotic ensemble of sub-shocks and oblique shocks may exist for a few days or weeks. This phenomenon may explain the brevity of the \ion{He}{II} $\lambda$4686 flash, its spherical symmetry (i.e., the \ion{He}{II} emission shows similar evolution in line strength and velocity when viewed from different directions), and its anti-correlation with the disappearance of 2--10 keV X-rays \citep{2011ApJ...740...80M,2012ApJ...746...73T}.
The relevant physics were summarized in \citet{2011ApJ...740...80M}; see also \citet{2002A&A...383..636P,2003ApJ...597..513S}.

Alternative explanations in which the \ion{He}{II} $\lambda 4686$ arises in the regular wind of the companion can be found in \citet{2004ApJ...612L.133S} and \citet{2006ApJ...652.1563S}.
Recently, \citet{2013MNRAS.436.3820M} presented results from SPH simulations, in which a complete WWC collapse is not required for the extra generation of soft X-rays at periastron passages. They proposed that the \ion{He}{II} emission is an effect of geometry. Soft X-rays from the WWC region can ionize the He$^{+}$ zone of $\eta$~Car's wind at 0.7--3~au when the WWC apex penetrates into it for orbital phases between  0.986 and 1.014. In this phase range the post-shock companion wind is in the radiative-cooling regime and is producing additional He$^{+}$ ionizing photons.
However, they stated that the sharp drop in \ion{He}{II} equivalent width at phase $\approx$1.0 may be due to a physical collapse of the WWC zone. They further suggested that the second, smaller emission peak could be due to a combination of instability in the WWC zone and increased optical depth in our line of sight caused by $\eta$~Car's wind. In fact, in order to explain the observed  \ion{He}{II} behavior at FOS4, this scenario very likely requires a collapse of the WWC. Detailed 3D modeling is required to determine if this is the case.

The greatest source of uncertainty in measuring the \ion{He}{II} emission strength is the determination of the underlying continuum level, see figure 3 in \citet{2006ApJ...640..474M}, figure 2 in \citet{2011ApJ...740...80M},  figure 2 in \citet{2012ApJ...746...73T}, and \citet{2014arXiv1411.0695D}. We followed \citet{2006ApJ...640..474M} and estimated the line strength by interpolating the continuum between 4601--4611\AA\ and 4739--4742~\AA. (The continuum ranges were slightly altered with respect to Martin et al.\ to take into account the strengthening of \ion{N}{II} lines near $\lambda$4600~\AA.) We then integrated the emission between 4673.5\AA\ and 4693.5~\AA\ ($-790$ to $+490$~km~s$^{-1}$). The resulting equivalent width is similar to EW2 in \citet{2014arXiv1411.0695D}. Figure \ref{fig:heIIstarfos4} shows the development of the equivalent width and peak velocity for \ion{He}{II} $\lambda$4686 during the 2014 events compared to previous cycles in 1998, 2003, and 2009, both for our direct line of sight and in the reflected spectra at FOS4. 
The values are listed in Tables \ref{table:heIIonstar} and \ref{table:heIIfos4}. 
Note that measurements of the equivalent widths in CHIRON data are higher than for STIS and UVES data. This arises, because the effective aperture of the CHIRON spectra is almost 3\arcsec\ and the spectra include faint emission lines originating from the ejecta around the central source \citep{2011PhDT........92M}.

\citet{2014arXiv1411.0695D} reported on the \ion{He}{II} $\lambda$4686 emission observed with STIS during the 2014 event. They concluded that $\eta$~Car's successive events differ in a progressive way. The \ion{He}{II} minimum may have been a quasi-eclipse by gas near the primary star, and the ``second $\lambda$4686 flash'' may have been stronger in 2014 because intervening densities were lower than in 2009.
In the discussion below, we include ground-based observations with UVES and CHIRON, which provide a higher time sampling for some parts of the event. UVES data also provide information of \ion{He}{II} from different stellar latitudes.

\subsubsection{The \ion{He}{II} maxima}  
\label{heII:1}

\begin{figure*}[t]
\centering 
\resizebox{\hsize}{!}{\includegraphics{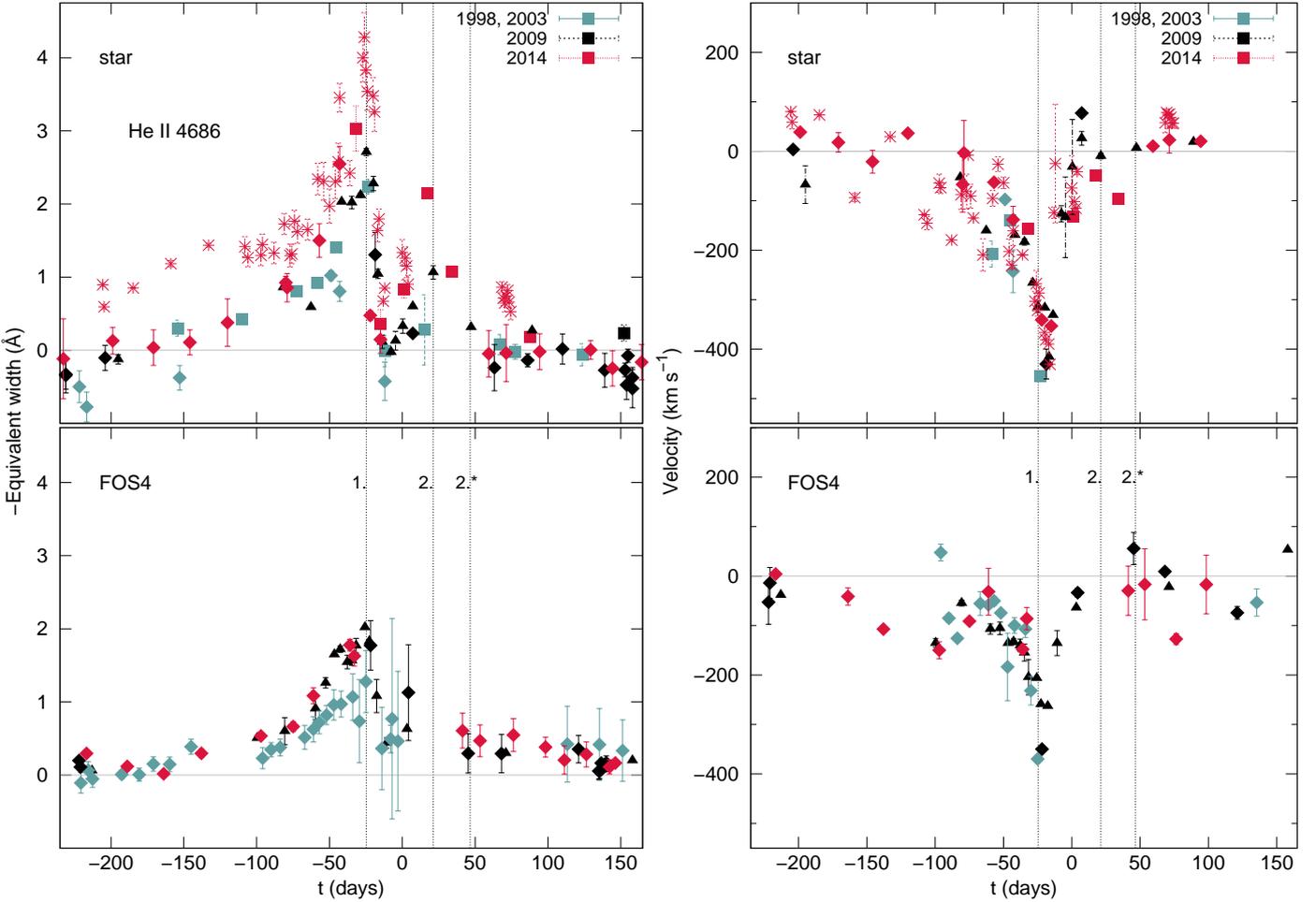}}
    \caption{Equivalent width and peak velocity of the \ion{He}{II} emission during the 1998--2014 events in direct view to the star and in the reflected spectra at FOS4 (squares: {\it HST\/} STIS, diamonds: VLT UVES, triangles: Gemini GMOS, asterisks: CTIO 1.5m CHIRON). Cycles are shifted by multiples of 2023~days. The values at FOS4 are corrected for the additional light travel time ($t_{\textnormal\scriptsize{FOS4}} = t+18$~d) and for the moving-mirror effect (${\Delta}\textnormal{v} = 100$~km~s$^{-1}$), see Section \ref{obs}.
The vertical lines indicate the timing of the approximate \ion{He}{II} peak emission. For the second \ion{He}{II} flare, we indicate two timings, labelled ``2.'' for the second peak in 2009 and 2014 and labelled ``2.$^*$'' for the assumed timing of the second peak in previous events based on the X-ray light curve and constraints from our spectra.}
     \label{fig:heIIstarfos4}
\end{figure*}

If the \ion{He}{II} emission is caused by a sensitive instability, then it is normal to expect large variations in equivalent widths and radial velocities between the observed periastron passages. Instead, the first occurrence of the $\lambda$4686 emission appears to reproduce the 2009 event better than for example the H$\alpha$ emission (with respect to the equivalent widths, see Section \ref{sec:halpha}). The second $\lambda$4686 emission flare, however, was notably stronger than for previous events.  
Like during previous events, the time of significant brightness of the \ion{He}{II} emission extended over $\sim$3~months. The line profile varied with significant changes on short timescales. It often resembled a single Gaussian, but on occasions it separated into two or more components, indicating a highly structured velocity- and time-variable WWC \citep{2014ATel.6334....1W}.

The time scale of the first (at $t\approx-25$~d) $\lambda$4686 flare in 2014 resembled the 2003 and 2009 occurrences. 
The flux grew concurrently with the decline of the 2--10~keV X-rays (see Figure \ref{fig:heIIxray} and Section \ref{sec:xrays}). The timing was consistent with a period of 2023~days. The maximum equivalent width may have been slightly larger than observed for the 2003 and 2009 events in direct view to the star. The apparent line flux, on the other hand, increased by a factor of $\sim$7 compared to 2003.  Of course this is partly due to the gradually decreasing extinction (which does not affect the \ion{He}{II} $\lambda$4686 equivalent widths, because the line emitting region is similar in size to the continuum emitting region). The velocities of the \ion{He}{II} emission are in good agreement with previous events. The line reached a maximum speed of $-430$~km~s$^{-1}$ on 2014 Jul 29. Only two days later a much smaller speed around $-100$~km~s$^{-1}$ is observed. The overall velocity range exceeds the maximum projected orbital velocity variation of any proposed 5.5~year orbit. Thus the velocities correspond likely to line-of-sight wind velocities and post-shock velocities, which can span a range of more than 400~km~s$^{-1}$ at various locations near periastron.

The evolution of the \ion{He}{II} $\lambda$4686 emission and velocity is very similar when viewed from different directions, i.e., in direct view of the star and reflected at FOS4, when the time delay is taken into account. 
Values for the equivalent widths and radial velocities are slightly smaller at FOS4 (Figure \ref{fig:heIIstarfos4}). At FOS4, the equivalent width followed closely the values of 2009, but was somewhat larger than in 2003. The same may be true for the values in our direct line of sight, but the sampling and the quality of the GMOS data are not good enough for a firm statement.  

We confirm the time delay of 18~days at FOS4 for the \ion{He}{II} emission \citep{2011ApJ...740...80M}. This supports the expected ${\Delta}t \approx 20$~days. The geometry of the reflection process of $\eta$~Car's stellar wind by the Homunculus nebula thus appears adequate.
Given the consistent time delay at FOS4, a good explanation is that the observed Doppler variations represent  ``global'' changes in outward wind and post-shock velocities that are not given a strong directionality by the secondary star as some authors have proposed \citep{2006ApJ...652.1563S,2007ApJ...660..669N}. A shock breakup model may conceivably act in a quasi-spherical way, with chaotic random velocity components during the critical time \citep{2011ApJ...740...80M}.  The predominance of negative Doppler velocities may indicate that the far side of the configuration is obscured by continuum Thomson scattering in the primary wind. 
 
 Perhaps the simplest conjecture is that \ion{He}{II} emission occurred in an outflow triggered by tidal forces, appearing more or less similar when viewed from most directions. Such an outflow can produce large varying Doppler blueshifts unrelated to the orbital velocity. (Redshifted material is not seen because it represents the obscured  far side of the outflow.) In this view, the emission may have occurred continuously during the interval when there were no hard X-rays, briefly disappearing when surrounding column densities were temporarily large enough to hide it.  
  This scenario is incomplete and obviously questionable, but it constitutes a useful initial hypothesis.

In \citet{2011ApJ...740...80M} we argued that, unlike the first \ion{He}{II} emission peak, the timing of the second \ion{He}{II} peak (at $t\approx20$~d in 2009 and 2014) may have differed in 2009 compared to previous events (compare the timings indicated with ``2.'' and ``2.${^*}$'' in Figure \ref{fig:heIIstarfos4}). We found that the 2003 event did not include a second \ion{He}{II} episode matching the one in 2009, but that in 2003 there was either no second episode or it was delayed. 
Contrary to \citet{2011ApJ...740...80M} and to \citet{2004ApJ...612L.133S}, \citet{2012ApJ...746...73T} stated that the second \ion{He}{II} emission peak occurred always at the same phase during all previous periastron passages. This discrepancy and its implications resulted in a great interest in the timing of the second \ion{He}{II} peak in 2014.

STIS observations on 2014 August 15 and CHIRON observations on 2014 August 13--18 suggested that the \ion{He}{II} $\lambda$4686 emission reappeared stronger, but likely at the same phase compared to 2009 (Figure \ref{fig:heIIstarfos4}). In 2014 August 31, a tremendously bright \ion{He}{II} $\lambda$4686 emission was observed in STIS spectra. The line's equivalent width was about twice as strong as the second peak in 2009 \citep{2014ATel.6448....1D}. It is unlikely that we missed a second \ion{He}{II} emission peak of similar strength in 2009 based on incomplete time coverage. Various observers covered the 2009 event quite well \citep{2011ApJ...740...80M,2012ApJ...746...73T}. The \ion{He}{II} strength rose and fell continuously, and there were no gaps long enough for a major spike. All three STIS data points during the second \ion{He}{II} emission peak in 2014 lie well above the interpolated Gemini and UVES measurements obtained in 2009 (Figure \ref{fig:heIIstarfos4}).
We discuss the significance of this second \ion{He}{II} episode in Section \ref{sec:xrays} below.

\subsubsection{The connection to the 2--10~keV X-rays}
\label{sec:xrays}

\begin{figure}
\centering
\resizebox{\hsize}{!}{\includegraphics{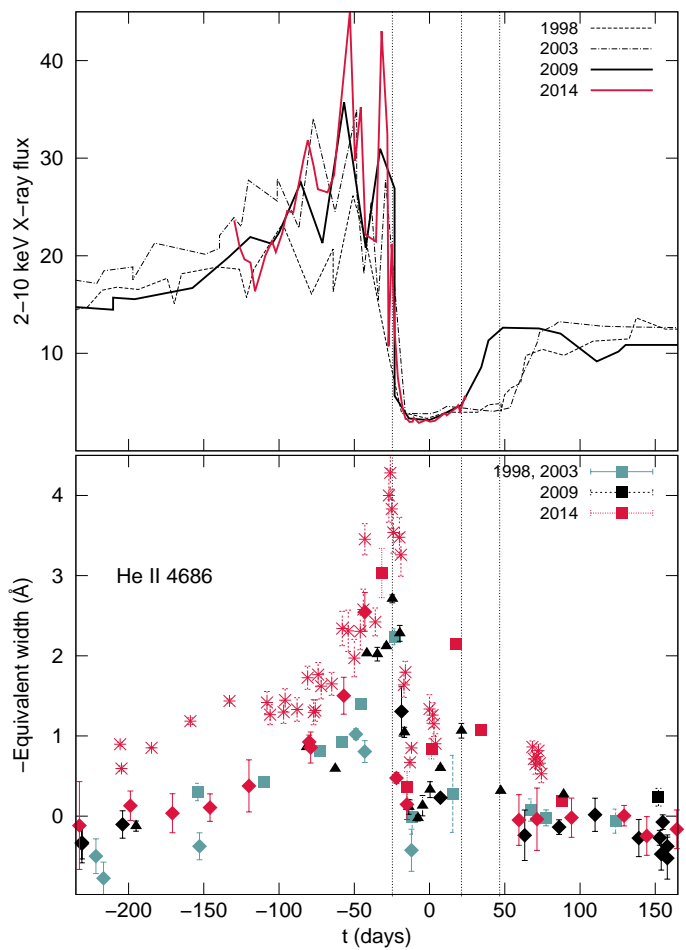}}
    \caption{Top: Schematic 2--10~keV X-ray light curve during the 1998--2014 events \citep{2014ATel.6453....1C}\protect\footnotemark[3]. Bottom: The \ion{He}{II} equivalent widths for the same time span. The vertical lines are to aid the comparison of the decline and rise of the X-rays with the \ion{He}{II} maxima. Unfortunately, we have no data at the expected later phase of the second \ion{He}{II} flare for events prior to 2009.}
     \label{fig:heIIxray}
\end{figure}

\citet{2002ASPC..262..267D} remarked that shock instabilities, rather than an eclipse scenario, can best  explain the rapid disappearance of $\eta$~Car's 2--10~keV X-rays during periastron passages.  
\citet{2003ApJ...597..513S} noted quantitative details and \citet{2006ApJ...652.1563S} suggested an instability involving accretion onto the companion and the shutdown of its wind. Other researchers adopted these ideas \citep{2008MNRAS.386.2330D,2009MNRAS.394.1758P,2011ApJ...740...80M,2012ApJ...746...73T}. 
\citet{2014ApJ...784..125H} suggested that the X-ray minimum observed during the 2003 event is the result of an eclipse of the WWC plasma during a  deep minimum followed by a collapse of the WWC activity seen during a shallow minimum. In 2009, the shallow minimum was not clearly seen and \citet{2014ApJ...795..119H} showed that the 15--25~keV X-ray emission observed during the 2009 event can be reproduced by WWC activity seen through the thick primary wind.

In this framework, two different causes but with similar large-scale consequences, have been proposed to explain the long X-ray minimum observed in previous events; shock instability and radiative inhibition.
1) A shock structure becomes unstable if radiative cooling exceeds expansion cooling due to thin shell instabilities \citep{1992ApJ...386..265S}. The slow primary's wind shock of $\eta$~Car is very unstable in this regard and the secondary's wind shock may become unstable near periastron, causing the entire shock structure to disintegrate on a timescale of  10--30~days \citep{2006ApJ...640..474M}.
2) Near periastron, soft X-rays from the shocked region and the radiation from the primary star can alter the ionization state of the companion's wind.  A higher degree of ionization weakens the line-driven acceleration, resulting in a slower wind speed (``radiative inhibition,'' \citealt{2009MNRAS.394.1758P,2011ApJ...726..105P}).  
 The consequence is that the balance of wind momenta is altered, pushing the shocks closer to the companion.  In an extreme case the primary wind can  entirely suppress the secondary wind \citep{2006ApJ...652.1563S}.     
 
The anticorrelation of the \ion{He}{II} emission and the X-rays (i.e., the \ion{He}{II} emission peaks at the times when the X-rays disappear and reappear) is consistent with a shock breakup model, and the second \ion{He}{II} $\lambda$4686 flash in 2009 strengthened the case \citep{2011ApJ...740...80M}. 
The \ion{He}{II} $\lambda$4686 feature appeared approximately when the hard X-rays peaked, it grew as they declined, and then ceased abruptly after a few weeks. The second \ion{He}{II} flare was qualitatively a reversal of the first episode. It occurred just as the X-rays reappeared and then declined concurrently with the growth of the X-rays. 
In the shock-instability scenario, we interpreted the second \ion{He}{II} peak as the re-formation of a large-scale shock structure when the relevant densities become sufficiently low for it to be quasi-stable. 
\footnotetext[3]{The X-ray light curve for previous cycles was retrieved at http://asd.gsfc.nasa.gov/Michael.Corcoran/eta\_car/etacar\_rxte\_lightcurve /index.html.}

In 2014, the timing of the 2--10~keV X-ray light curve with its peak, minimum, and reappearance \citep{2014ATel.6453....1C,2014ATel.6357....1C} and its relation to the \ion{He}{II} emission was similar to the 2009 event. That is, also for the 2014 event, the second \ion{He}{II} flare occurred when the X-rays reappeared. \footnote{ Unfortunately, only the first part of the X-ray egress out of the 2014 minimum, but not the full recovery has been published at the time of submission.} The similar timing of the X-ray and \ion{He}{II} emission during the 2014 and 2009 events does not contradict that there is a change in the system. The motion of the secondary star at periastron is very fast and, if gas densities are somewhat different, this would alter the critical times of observables by only a few days. The X-ray minima in 2009 and 2014, however, were much shorter than the ones observed in 1998 and 2003. This cannot be explained with an unchanged system and a simple wind occultation scenario. Unfortunately, we have no data points during the expected (delayed) time of the second flare for the events in 1998 and 2003 to prove the correlation between the second \ion{He}{II} flare  and the reappearance of the X-rays.

The second \ion{He}{II} $\lambda$4686 flash in 2014 was much stronger than in 2009 (Figures \ref{fig:heIIstarfos4} and \ref{fig:heIIxray}). This together with the short X-ray minimum (without a shallow phase, see \citealt{2014ApJ...784..125H,2014ApJ...795..119H}) may indicate that a complete shock breakup -- needed to explain the weak second \ion{He}{II} flare and the long X-ray minimum in previous events (see \citealt{2009ApJ...701L..59K} and \citealt{2006ApJ...652.1563S}) -- may not have occurred in 2014. Instead, the 2014 event resembled more an occultation by the primary wind. The \ion{He}{II} emission may have disappeared temporarily when the column densities in our line of sight to the emitting region as the secondary passed behind the primary's wind were large enough. {}

\section{Discussion}
\label{discussion}

Several observables indicate that physical parameters of the stellar system have changed: 
1) The continuous weakening of the H$\alpha$ and \ion{Fe}{II} lines from the primary wind. 
2) The tremendous increase in flux of the \ion{He}{I} and \ion{N}{II} emission lines over the last 3 cycles, and especially during the 2014 event. 
3) The much stronger line flux of the \ion{He}{II} emission in 2014 compared to 2009 and its stronger reappearance after the X-ray minimum in 2014. 

The simplest interpretation for the decrease of the broad primary wind emission features is a decrease in mass-loss rate. The simultaneous increase in the higher ionization/excitation species, \ion{He}{I} and \ion{N}{II}, indicates that the companion's effective far-UV radiation is much more efficient than it was in previous cycles -- consistent with lower gas densities. As the wind density decreases it is conceivable that the primary star may eventually become ``hot'' enough to excite these lines in addition to the secondary. 

The SPH simulations by \citet{2013MNRAS.436.3820M} showed that a factor of two or more decrease in mass loss alters significantly the time-dependent 3D density, temperature, and velocity structure of $\eta$~Car's spatially-extended primary wind and WWC zones. They showed that a lower mass-loss rate would lead to weaker emission and absorption of the broad wind-emission features of H$\alpha$ and \ion{Fe}{II} in our line of sight and at the stellar poles. Emission lines of \ion{He}{I} would strengthen. Based on their simulations, a drop in mass loss by a factor of $\lesssim2$ seems to reasonably fit our observations, but further modelling is required.

The spectrum at FOS4 shows less significant changes in the broad wind emission lines than the spectrum in our direct line of sight. The H$\alpha$ equivalent widths display a very similar behavior during the 2003, 2009, and 2014 events. The H$\alpha$ P~Cyg absorption strength decreased somewhat. The broad \ion{Fe}{II} emission features decreased in strengths -- albeit less than in our direct line of sight. On the other hand, the \ion{He}{II} equivalent widths are higher in 2009 and 2014 than in 2003 in accordance with the observations in our direct line of sight.

\citet{2005ASPC..332..169O} suggested that the rapid (near-critical) stellar rotation of $\eta$~Car induces an equatorial gravity darkening \citep{1996ApJ...472L.115O}.   
It may be that the companion star spins up $\eta$~Car incrementally by angular momentum transfer between the two stars at periastron  through tidal torques  \citep{1997NewA....2..387D}. Its rotational velocity is approaching its critical velocity, leading to lower effective gravity at the equator, and thus reducing the mass-loss rate there. Less changes in the broad wind emission lines observed in the polar spectrum at FOS4 are thus not surprising.   In any case, the observed changes in \ion{Fe}{II} and helium emission  at FOS4 are sufficient to confirm a secular trend in $\eta$~Car's wind -- independent of the direct-view data.  The quantitative rates require models far beyond the scope of this paper.

The polar view (FOS4) is  valuable for the specific purposes noted earlier, but it is not a  {\it typical\/} view.  If, for  example, it applies within 30\degree\ of each pole, then that would include only 13\% of possible viewing directions. Eta~Car's secular changes appear larger for lower latitudes,  i.e., most directions in space. We mention this in order to emphasize that the direction-averaged mass-loss rate has always been unclear.

Some details occur at nearly the same phase in each event, for instance the rapid decline of \ion{He}{II} $\lambda$4686 near $t \approx -17$~d. Rapid orbital motion can largely explain this fact almost independent of the flow parameters. Consider, for example, an orbit with eccentricity $\epsilon = 0.85$. During a 40-day interval around periastron, the system sweeps through almost 4\degree\ of longitude per day. Tidal effects change by about 6 percent per day, the colliding-wind instability parameters likewise vary rapidly, and the same is true for possible eclipse parameters (see figure 10 in \citealt{2011ApJ...740...80M}).   
Thus an ambient density decrease of 30 percent between events would be compensated by orbital motion within 
a few days.  Moreover, the primary mass outflow during $t \approx -45$~d to $-10$~d is very likely enhanced by tidal 
and radiative effects which are unrelated to the current average mass-loss rate \citep{2006ApJ...640..474M,2011ApJ...740...80M}.    
For these reasons, the timing of events does not contradict the secular trend in wind density. If we avoid definitions that are adjusted {\it ex post facto\/} to optimize the correlations, then, objectively, the data allow variations of 3--10~days between events. 
A more eccentric orbit with $\epsilon \approx 0.90$, favored by most authors in recent years, entails even faster rates of change.

In previous events, the combined \ion{He}{II} $\lambda$4686 and X-ray data, i.e., the extended X-ray minima and 
the association of the \ion{He}{II} flares with the appearance and disappearance of the X-rays, were persuasively consistent with a shock breakup scenario and mass accretion onto the secondary. 
The stronger second \ion{He}{II} flare in 2014 indicates less mass accretion onto the secondary. Conceivably, the shock structure may even have not been completely destroyed in 2014 and the X-ray and \ion{He}{II} behavior may be explained by an eclipse of the secondary by the dense primary wind. 

\section{Conclusion}
\label{conclusion}

It is conceivable that 2009.1 was the {\it last\/} $\eta$~Car event of the classic type, and 2014.6 began a new regime. An approximate historical time line may look something like this:

\begin{enumerate}
   \item 1900--1948, $\eta$~Car's primary wind was so dense
      that it totally suppressed the secondary wind.
      Therefore a small fraction of the outflowing
      gas accreted onto the secondary star, making
      it look like a cooler star.  No helium-ionizing
      far-UV photons escaped from the secondary star and thus no high-excitation lines were observed \citep{2008AJ....135.1249H}. \citet{2015MNRAS.447.2445C} showed that for mass-loss rates of more than a factor of 2--4 than today, accretion may not have been needed to explain the absence of \ion{He}{I} lines.
   
   \item During the 1940's, the mass-loss rate decreased
      enough so that the secondary star could develop a
      normal hot-star wind \citep{2008AJ....135.1249H}.  Accretion
      stopped except near periastron and the familiar
      wind-wind shocks formed.
   
   \item 1948--2009, the wind-wind shocks were disrupted
      near each periastron passage. That allowed temporary
      accretion onto the secondary star \citep{2003ApJ...597..513S,2005ApJ...635..540S,2006ApJ...652.1563S}. Less far-UV radiation
      during each event resulted in the high-excitation lines to become temporarily weaker.
   
   \item 2014, the primary wind density has fallen
      low enough so that full accretion no longer occurs.
      Therefore, much more UV emission than ever before can escape from the secondary and penetrate the primary wind.
      This would help to explain the strong \ion{He}{I} and \ion{N}{II}
      UV-excited lines. In addition, the primary star may be have become hot enough to contribute additional UV radiation.

\end{enumerate}

This rough history of events leaves unaccounted some observed features. 
For example, the high-excitation lines from the Weigelt knots are associated with the secondary star and their disappearance during the events are interpreted as a shut-down of the UV radiation from the secondary. Still, they disappeared in 2014. However, they are located near the bipolar midplane (probably the same as the orbit plane), at longitudes roughly opposite to the secondary's periastron. Therefore, during the events, the primary wind very likely shields them from the secondary star's UV radiation.
Spectra at FOS4 have not experienced the same dramatic changes in the broad wind emission lines as observations in our line of sight. This may be explained by a latitude-dependent wind structure of $\eta$~Car.

With the ground- and space-based observations obtained during the 2014 event we aimed to further investigate the long-term changes in $\eta$~Car and the impact of the periastron passages on its long-term evolution.
On top of the orbital modulations, the broad primary wind features have decreased in strength and the higher-ionization \ion{He}{I} and \ion{N}{II} emission lines have grown in strength over the past 10 years. This marks an undeniable change in the physics of the region. A decreased mass-loss rate could explain the weakened low-excitation wind features, the strengthening of higher ionization/excitation lines, and the \ion{He}{II} and X-ray behavior. 
A decrease in the primary wind density would not only allow the intense UV radiation from the stars to ionize/excite more material, but also to more easily destroy dust and thus decrease the extinction in our line of sight.

\begin{acknowledgements} We are grateful to the STScI staff members for help in scheduling the intricate STIS observations. We thank the Paranal Observatory for conducting the UVES observations. Stony Brook's continuing involvement in the SMARTS Consortium is made possible by generous support from the Provost of Stony Brook University.
AM was co-funded under the Marie Curie Actions of the European Commission (FP7-COFUND). We thank Jose Groh for valuable discussions and comments on the paper draft. We thank the anonymous referee for the insistence to distinguish more clearly between the two lines of sights described in this paper.
\end{acknowledgements}

\bibliographystyle{aa}

\begin{longtable}{lllllcc}
\caption{\label{table:journal} Journal of observations.}\\
\hline\hline
Instrument & Position & Date & MJD  & Phase  & Wavelength$^a$ &  Exposure Time$^b$   \\ 
& & & (days)  &  & (\AA)  & (s) \\ 
\hline
\endfirsthead
\caption{continued.}\\
\hline\hline
Instrument & Location & Date & MJD  & Phase  & Wavelength$^a$ &  Exposure Time$^b$   \\ 
& & & (days)  &  & (\AA)  & (s) \\ 
\hline
\endhead
\hline
\endfoot
STIS/CCD&	Star	&	2013-09-14	&	56549.2	&	2.835	&	6488--7051	&	2$\times$0.2; 2$\times$2; 2$\times$15	\\
STIS/CCD	&	Star	&	2013-09-14	&	56549.2	&	2.835	&	5454--6018		&	2$\times$4 \\
STIS/CCD&	Star	&	2013-09-14	&	56549.2	&	2.835	&	3796--4077		&	2$\times$27	\\
STIS/CCD	&	Star	&	2013-09-14	&	56549.2	&	2.835	&	4561--4841	&	2$\times$30; 8 \\
STIS/CCD	&	Star	&	2013-09-14	&	56549.3	&	2.835	&	4820--5099	&	2$\times$2.4; 2$\times$20	\\
STIS/CCD	&	Star	&	2013-09-14	&	56549.3	&	2.835	&	2913--5702	&	4	\\

STIS/CCD&	Star	&	2014-07-13	&	56851.2	&	2.984	&	6487--7051	&	0.2; 2	\\
STIS/CCD	&	Star	&	2014-07-13	&	56851.2	&	2.984	&	2912--5702	&	3	\\
STIS/CCD	&	Star	&	2014-07-13	&	56851.2	&	2.984	&	4561--4841	&	25.5		\\
STIS/CCD	&	Star	&	2014-07-13	&	56851.2	&	2.984	&	3796--4077	&	24	\\

STIS/CCD	&	Star	&	2014-07-30	&	56868.1	&	2.993	&	6487--7051	&	0.2; 2	\\
STIS/CCD&	Star	&	2014-07-30	&	56868.1	&	2.993	&	2913--5702	&	3	\\
STIS/CCD&	Star	&	2014-07-30	&	56868.1	&	2.993	&	4561--4841	&	25.5		\\
STIS/CCD	&	Star	&	2014-07-30	&	56868.1	&	2.993	&	3796--4077	&	24	\\

STIS/CCD	&	Star	&	2014-08-15	&	56868.1	&	3.001	&	6487--7051	&	0.2; 2	\\
STIS/CCD&	Star	&	2014-08-15	&	56868.1	&	3.001	&	2913--5702	&	3	\\
STIS/CCD&	Star	&	2014-08-15	&	56868.1	&	3.001	&	4561--4841	&	25.5		\\
STIS/CCD	&	Star	&	2014-08-15	&	56868.1	&	3.001	&	3796--4077	&	24	\\

STIS/CCD	&	Star	&	2014-08-31	&	56900.4	& 3.009	&	6487--7051	&	0.2; 2     \\	
STIS/CCD	&	Star	&	2014-08-31	&	56900.4	& 3.009	&	2913--5702	&	3	\\	
STIS/CCD	&	Star	&	2014-08-31	&	56900.4	& 3.009	&	4561--4841	&	25.5		\\	
STIS/CCD	&	Star	&	2014-08-31	&	56900.4	& 3.009	&	3796--4077	&	24	\\	

STIS/CCD	&	Star	&	2014-09-17	&	56917.0	&	3.017	&	6487--7051  	& 0.2; 2	\\	
STIS/CCD	&	Star	&	2014-09-17	&	56917.0	&	3.017	&	2913--5702 	&	3	\\	
STIS/CCD	&	Star	&	2014-09-17	&	56917.1	&	3.017	&	4561--4841  	&	25.5	\\	
STIS/CCD	&	Star	&	2014-09-17	&	56917.1	&	3.017	&	3796--4077  	&	24	\\	

STIS/CCD	&	Star	&	2014-11-09	&	56970.9	&	3.043	&	6487--7051  	& 0.2; 2	\\	
STIS/CCD	&	Star	&	2014-11-09	&	56970.9	&	3.043	&	2913--5702 	&	3	\\	
STIS/CCD	&	Star	&	2014-11-09	&	56970.9	&	3.043	&	4561--4841  	&	25.5	\\	
STIS/CCD	&	Star	&	2014-11-09	&	56970.9	&	3.043	&	3796--4077  	&	24	\\

\hline

UVES	&	Star	&	2012-04-03	&	56020.1	&	2.586	&	DIC1		&	250, 150, 100, 50 (blue); 4$\times$20, 4$\times$5, 3$\times$1, 0.7 (red) \\
UVES	&	Star	&	2012-04-03	&	56020.1	&	2.586	&	DIC2		&	2$\times$20, 2$\times$10, 2$\times$5 (blue); 50, 20, 10 (red)	\\

UVES	&	FOS4	&	2012-04-04	&	56021.1	&	2.587	&	DIC1		&	600, 300, 100, 10 (blue); 2$\times$200, 3$\times$60, 2$\times$20, 10 (red)	\\
UVES	&	FOS4	&	2012-04-04	&	56021.1	&	2.587	&	DIC2		&	500, 300, 90 (blue); 500, 300, 90 (red)	\\

UVES	&	Star	&	2013-11-24	&	56620.3	&	2.884	&	DIC2		&	20, 10 (blue); 10, 5 (red)	\\
UVES	&	FOS4	&	2013-11-24	&	56620.3	&	2.884	&	DIC2		&	250 (blue); 200 (red)	\\

UVES	&	Star	&	2013-11-25	&	56621.3	&	2.885	&	DIC1		&	100, 50 (blue); 2$\times$15, 2$\times$0.6 (red)	\\
UVES	&	FOS4	&	2013-11-25	&	56621.3	&	2.885	&	DIC1		&	600, 100	 (blue); 2$\times$200, 2$\times$50 (red)	\\

UVES	&	Star	&	2013-12-24	&	56650.3	&	2.899	&	DIC1		&	100, 2$\times$50	, 10 (blue); 2$\times$15, 2$\times$5, 4$\times$0.5 (red)	\\
UVES	&	Star	&	2013-12-24	&	56650.3	&	2.899	&	DIC2		&	20, 10 (blue); 10, 5	 (red)\\
UVES	&	FOS4	&	2013-12-24	&	56650.3	&	2.899	&	DIC1		&	600, 100	 (blue); 2$\times$200, 2$\times$50	 (red)\\
UVES	&	FOS4	&	2013-12-24	&	56650.3	&	2.899	&	DIC2 	&	250 (blue); 200 (red)	\\

UVES	&	Star	&	2014-01-27	&	56684.2	&	2.916	&	DIC1		&	100, 50 (blue); 2$\times$15, 2$\times$0.5 (red)	\\
UVES	&	Star	&	2014-01-27	&	56684.2	&	2.916	&	DIC2		&	20, 10 (blue); 10, 5 (red) \\
UVES	&	FOS4	&	2014-01-27	&	56684.2	&	2.916	&	DIC1		&	600, 100	 (blue); 2$\times$200, 2$\times$50 (red)\\
UVES	&	FOS4	&	2014-01-27	&	56684.2	&	2.916	&	DIC2		&	250 (blue); 200	 (red)\\

UVES	&	Star	&	2014-02-24	&	56712.2	&	2.930	&	DIC1		&	100, 50 (blue); 2$\times$15, 0.5	 (red)\\
UVES	&	Star	&	2014-02-24	&	56712.2	&	2.930	&	DIC2		&	20, 10 (blue); 10, 5	 (red)\\
UVES	&	FOS4	&	2014-02-24	&	56712.2	&	2.930	&	DIC1		&	600, 100	 (blue); 2$\times$200, 2$\times$50	 (red)\\
UVES	&	FOS4	&	2014-02-24	&	56712.2	&	2.930	&	DIC2		&	250 (blue); 200 (red)	\\

UVES	&	Star	&	2014-03-21	&	56737.1	&	2.942	&	DIC1		&	100, 50 (blue); 2$\times$15, 2$\times$0.5	 (red)\\
UVES	&	Star	&	2014-03-21	&	56737.1	&	2.942	&	DIC2		&	20, 10 (blue); 10, 5	 (red)\\
UVES	&	FOS4	&	2014-03-21	&	56737.2	&	2.942	&	DIC1		&	600, 100 (blue); 2$\times$200, 2$\times$50 (red)	\\
UVES	&	FOS4	&	2014-03-21	&	56737.2	&	2.942	&	DIC2		&	250 (blue); 200	 (red)\\

UVES	&	Star	&	2014-04-16	&	56763.0	&	2.955	&	DIC1		&	100, 50 (blue); 2$\times$15, 2$\times$0.5 (red)	\\
UVES	&	Star	&	2014-04-16	&	56763.0	&	2.955	&	DIC2		&	20, 10 (blue); 10, 5 (red)\\
UVES	&	FOS4	&	2014-04-16	&	56763.1	&	2.955	&	DIC1		&	400 (blue); 2$\times$170 (red)	\\
UVES	&	FOS4	&	2014-04-16	&	56763.1	&	2.955	&	DIC2		&	200 (blue); 200 (red)	\\

UVES	&	Star	&	2014-05-26	&	56803.0	&	2.975	&	DIC1		&	3$\times$100, 3$\times$50 (blue); 5$\times$15, 6$\times$0.5	 (red)\\
UVES	&	Star	&	2014-05-26	&	56803.0	&	2.975	&	DIC2		&	3$\times$20, 3$\times$10 (blue); 3$\times$10, 3$\times$5  (red)\\
UVES	&	FOS4	&	2014-05-26	&	56803.0	&	2.975	&	DIC1		&	2$\times$400	 (blue); 4$\times$170 (red)	\\

UVES	&	Star	&	2014-05-27	&	56804.0	&	2.976	&	DIC1		&	100, 50 (blue); 2$\times$15, 2$\times$0.5	 (red)\\
UVES	&	Star	&	2014-05-27	&	56804.0	&	2.976	&	DIC2		&	20, 10 (blue); 10, 5  (red)\\
UVES	&	FOS4	&	2014-05-27	&	56804.0	&	2.976	&	DIC1		&	400 (blue); 2$\times$170 (red)	\\
UVES	&	FOS4	&	2014-05-27	&	56804.0	&	2.976	&	DIC2		&	200 (blue); 200 (red)	\\

UVES	&	Star	&	2014-06-18	&	56826.0	&	2.987	&	DIC1		&	100, 50 (blue); 2$\times$15, 2$\times$0.5, 2$\times$0.2	 (red)\\
UVES	&	Star	&	2014-06-18	&	56826.0	&	2.987	&	DIC2		&	20, 10 (blue); 10, 5  (red)\\
UVES	&	FOS4	&	2014-06-18	&	56826.0	&	2.987	&	DIC1		&	400 (blue); 2$\times$170, 2$\times$30	 (red)\\
UVES	&	FOS4	&	2014-06-18	&	56826.1	&	2.987	&	DIC2		&	200 (blue); 200	 (red)\\

UVES	&	Star	&	2014-07-02	&	56840.0	&	2.994	&	DIC1		&	100, 50 (blue); 2$\times$15, 2$\times$0.5	 (red)\\
UVES	&	Star	&	2014-07-02	&	56840.0	&	2.994	&	DIC2		&	20, 10 (blue); 10, 5  (red)	\\
UVES	&	FOS4	&	2014-07-02	&	56840.0	&	2.994	&	DIC1		&	400 (blue); 2$\times$170	 (red)\\
UVES	&	FOS4	&	2014-07-02	&	56840.0	&	2.994	&	DIC2		&	200 (blue); 200	 (red)\\

UVES	&	Star &  2014-07-23 &  56861.0 &  2.989 &  DIC1	 &  100, 50 (blue); 2$\times$15 , 2$\times$1   (red)\\ 
UVES	&	Star &  2014-07-23 &  56861.0 &  2.989 &  DIC2	 &  20, 10 (blue); 10, 5  (red) \\ 
UVES	&	FOS4	 &  2014-07-23 &  56861.0 &  2.989 &  DIC1	&  400 (blue); 2$\times$170, 2$\times$10  (red) \\ 

UVES	&	FOS4 &  2014-07-27 &  56865.0 &  2.991 &  DIC1	 &  2$\times$400, 200 (blue); 4$\times$170, 2$\times$85   (red)\\ 
UVES	&	FOS4 &  2014-07-27 &  56865.0 &  2.991 &  DIC2	 &  2$\times$200 (blue); 200, 80   (red)\\ 

UVES	&	Star &  2014-07-30 &  56868.0 &  2.993 &  DIC1	 &  100, 50 (blue); 2$\times$15, 2$\times$1  (red) \\ 
UVES	&	Star &  2014-07-30 &  56868.0 &  2.993 &  DIC2	 &  20, 10 (blue); 10, 5 (red)  \\ 
UVES	&	FOS4 &  2014-07-30 &  56868.0 &  2.993 &  DIC1	 &  400 (blue); 2$\times$170   (red)\\ 
UVES	&	FOS4 &  2014-07-30 &  56868.0 &  2.993 &  DIC2	 &  200 (blue); 200  (red) \\ 

UVES	&	Star &  2014-10-12 &  56942.4 &  3.029 &  DIC1	 &  100, 50 (blue); 2$\times$15, 2$\times$1  (red) \\ 
UVES	&	Star &  2014-10-12 &  56942.4 &  3.029 &  DIC2	 &  20, 10 (blue); 10, 5 (red)  \\ 
UVES	&	FOS4 &  2014-10-12 &  56942.4 &  3.029 &  DIC1	 &  400 (blue); 2$\times$170   (red)\\ 
UVES	&	FOS4 &  2014-10-12 &  56942.4 &  3.029 &  DIC2	 &  200 (blue); 200  (red) \\ 

UVES	&	Star &  2014-10-24 &  56954.4 &  3.035 &  DIC1	 &  100, 50 (blue); 2$\times$15, 2$\times$1  (red) \\ 
UVES	&	Star &  2014-10-24 &  56954.4 &  3.035 &  DIC2	 &  20, 10 (blue); 10, 5 (red)  \\ 
UVES	&	FOS4 &  2014-10-24 &  56954.4 &  3.035 &  DIC1	 &  400 (blue); 2$\times$170   (red)\\ 
UVES	&	FOS4 &  2014-10-24 &  56954.4 &  3.035 &  DIC2	 &  200 (blue); 200  (red) \\ 

UVES	&	Star &  2014-11-16 &  56977.3 &  3.047 &  DIC1	 &  100, 50 (blue); 2$\times$15, 2$\times$1  (red) \\ 
UVES	&	Star &  2014-11-16 &  56977.3 &  3.047 &  DIC2	 &  20, 10 (blue); 10, 5 (red)  \\ 
UVES	&	FOS4 &  2014-11-16 &  56977.3 &  3.047 &  DIC1	 &  400 (blue); 2$\times$170   (red)\\ 
UVES	&	FOS4 &  2014-11-16 &  56977.3 &  3.047 &  DIC2	 &  200 (blue); 200  (red) \\ 

UVES	&	FOS4 &  2014-12-08 &  56999.3 &  3.058 &  DIC1	 &  400 (blue); 2$\times$170   (red)\\ 
UVES	&	FOS4 &  2014-12-08 &  56999.3 &  3.058 &  DIC2	 &  200 (blue); 200  (red) \\ 

UVES	&	Star &  2014-12-21 &  57012.3 &  3.064 &  DIC1	 &  100, 50 (blue); 2$\times$15, 2$\times$1  (red) \\ 
UVES	&	Star &  2014-12-21 &  57012.3 &  3.064 &  DIC2	 &  20, 10 (blue); 10, 5 (red)  \\ 
UVES	&	FOS4 &  2014-12-21 &  57012.3 &  3.064 &  DIC1	 &  400 (blue); 2$\times$170   (red)\\ 
UVES	&	FOS4 &  2014-12-21 &  57012.3 &  3.064 &  DIC2	 &  200 (blue); 200  (red) \\ 

UVES	&	Star &  2015-01-05 &  57027.3 &  3.071 &  DIC1	 &  100, 50 (blue); 2$\times$15, 2$\times$1  (red) \\ 
UVES	&	Star &  2015-01-05 &  57027.3 &  3.071 &  DIC2	 &  20, 10 (blue); 10, 5 (red)  \\ 
UVES	&	FOS4 &  2015-01-05 &  57027.3 &  3.071 &  DIC1	 &  400 (blue); 2$\times$170   (red)\\ 
UVES	&	FOS4 &  2015-01-05 &  57027.3 &  3.071 &  DIC2	 &  200 (blue); 200  (red) \\ 

UVES	&	FOS4 &  2015-01-21 &  57043.4 &  3.079 &  DIC2	 &  200 (blue); 200  (red) \\ 

UVES	&	Star &  2015-01-25 &  57047.3 &  3.081 &  DIC1	 &  100, 50 (blue); 2$\times$15, 2$\times$1  (red) \\ 
UVES	&	Star &  2015-01-25 &  57047.3 &  3.081 &  DIC2	 &  20, 10 (blue); 10, 5 (red)  \\ 
UVES	&	FOS4 &  2015-01-25 &  57047.3 &  3.081 &  DIC1	 &  400 (blue); 2$\times$170   (red)\\ 
UVES	&	FOS4 &  2015-01-25 &  57047.3 &  3.081 &  DIC2	 &  200 (blue); 200  (red) \\ 

\hline \hline
\multicolumn{7}{l}{\tablefootmark{(a)} DIC1: $\lambda$3044--3916, $\lambda$4726--5803, $\lambda$5762--6835; DIC2: $\lambda$3732--4999, $\lambda$6649--8544, $\lambda$8523--10426.} \\
\multicolumn{7}{l}{\tablefootmark{(b)} STIS/CCD slit: 52\arcsec$\times$0\farcs1. UVES slit widths: 0\farcs4 (blue arm) and 0\farcs3 (red arm). } \\
\end{longtable}



\begin{longtable}{llcccccrr}
\caption{\label{table:heIIonstar} \ion{He}{II} equivalent width and velocity in direct view of the star. Velocity is the peak velocity.}\\
\hline\hline
Position &	Instrument	& Date & MJD  & Phase  & $-$EW & $\Delta$EW   & Vel$^a$ & $\Delta$Vel  \\ 
 &	& & (days)  &  & (\AA)& (\AA) & (km~s$^{-1}$)& (km~s$^{-1}$) \\ 
\hline
\endfirsthead
\caption{continued.}\\
\hline\hline
Location &	Instrument	& Date & MJD  & Phase  & $-$EW & $\Delta$EW   & Vel$^a$ & $\Delta$Vel    \\ 
& & & (days)  &  & (\AA)& (\AA) & (km~s$^{-1}$)& (km~s$^{-1}$) \\ 
\hline
\endhead
\hline
\endfoot
Star & STIS/CCD & 1998-03-19 & 50891.4 & 0.038 & 0.160 & 0.050 &  \\ 
Star & STIS/CCD & 1999-02-21 & 51230.5 & 0.206 & -0.080 & 0.060 &  \\ 
Star & STIS/CCD & 2000-03-20 & 51623.8 & 0.400 & -0.120 & 0.070 &  \\ 
Star & STIS/CCD & 2001-04-17 & 52016.8 & 0.595 & -0.030 & 0.060 &  \\ 
Star & STIS/CCD & 2002-01-20 & 52294.0 & 0.732 & 0.010 & 0.050 &  \\ 
Star & STIS/CCD & 2002-07-04 & 52459.5 & 0.813 & 0.140 & 0.040 &  \\ 
Star & STIS/CCD & 2003-02-13 & 52683.1 & 0.924 & 0.440 & 0.040 &  \\ 
Star & STIS/CCD & 2003-03-29 & 52727.3 & 0.946 & 0.560 & 0.040 &  \\ 
Star & STIS/CCD & 2003-05-05 & 52764.3 & 0.964 & 1.010 & 0.040 &  \\ 
Star & STIS/CCD & 2003-05-19 & 52778.5 & 0.971 & 1.110 & 0.080 & -208 & 26 \\ 
Star & STIS/CCD & 2003-06-01 & 52791.7 & 0.978 & 1.630 & 0.080 & -140 & 3 \\ 
Star & STIS/CCD & 2003-06-23 & 52813.8 & 0.989 & 2.590 & 0.060 & -455 & 1 \\ 
Star & STIS/CCD & 2003-07-05 & 52825.4 & 0.994 & 0.160 & 0.060 &  \\ 
Star & STIS/CCD & 2003-08-01 & 52852.4 & 1.008 & 0.150 & 0.080 &  \\ 
Star & STIS/CCD & 2003-09-22 & 52904.3 & 1.033 & 0.150 & 0.010 &  \\ 
Star & STIS/CCD & 2003-11-17 & 52960.6 & 1.061 & 0.150 & 0.040 &  \\ 
Star & STIS/CCD & 2004-03-07 & 53071.2 & 1.116 & 0.170 & 0.040 &  \\ 
Star & STIS/CCD & 2009-06-30 & 55012.1 & 2.075 & 0.236 & 0.110 &  \\ 
Star & STIS/CCD & 2009-08-19 & 55062.0 & 2.100 & 0.166 & 0.057 &  \\ 
Star & STIS/CCD & 2009-12-06 & 55171.6 & 2.154 & 0.134 & 0.020 &  \\ 
Star & STIS/CCD & 2010-03-03 & 55258.6 & 2.197 & 0.121 & 0.058 &  \\ 
Star & STIS/CCD & 2010-08-20 & 55428.3 & 2.281 & 0.035 & 0.117 &  \\ 
Star & STIS/CCD & 2010-10-26 & 55495.1 & 2.314 & -0.018 & 0.063 &  \\ 
Star & STIS/CCD & 2011-11-20 & 55885.6 & 2.507 & 0.129 & 0.011 &  \\ 
Star & STIS/CCD & 2012-10-18 & 56218.5 & 2.672 & 0.099 & 0.027 &  \\
Star & STIS/CCD & 2013-09-14 & 56549.2 & 2.835 & 0.276 & 0.061 &  \\ 
Star & STIS/CCD & 2014-07-13 & 56851.2 & 2.984 & 3.032 & 0.308 & -157 & 2 \\
Star & STIS/CCD & 2014-07-30 & 56868.1 & 2.993 & 0.366 & 0.188 &  \\ 
Star & STIS/CCD & 2014-08-15 & 56884.3 & 3.001 & 0.838 & 0.124 & -132 & 4 \\ 
Star & STIS/CCD & 2014-08-31	& 56900.4 &	3.009 & 	2.152 &	0.030 &	-49 & 7 \\
Star & STIS/CCD & 2014-09-17	& 56917.1 &	3.017 & 1.073 &	0.022 &	 -97 	&	2	\\	
Star & STIS/CCD & 2014-11-09	& 56970.9 &	3.043 & 0.151 &	0.022 &	 	&	\\	
		\hline 
Star &	UVES	&	2002-12-07	&		52615.3	&	0.890	&	-0.500	&	0.217 \\
Star &	UVES	&	2002-12-12	&		52620.3	&	0.893	&	-0.775	&	0.202 \\
Star &	UVES	&	2003-02-14	&		52684.1	&	0.924	&	-0.376	&	0.168 \\
Star &	UVES	&	2003-05-29	&		52788.1	&	0.976	&	1.022	&	0.067	&	-97 &	2 \\
Star &	UVES	&	2003-06-03	&		52794.0	&	0.979	&	0.806	&	0.136	&	-242 &	43	\\
Star &	UVES	&	2003-07-05	&		52825.0	&	0.994	&	-0.426	&	0.263 \\
Star &	UVES	&	2004-02-20	&		53055.1	&	1.108	&	1.283	&	0.054 \\
Star &	UVES	&	2005-02-12	&		53413.4	&	1.285	&	-0.790	&	0.243 \\
Star &	UVES	&	2005-03-19	&		53448.1	&	1.302	&	-0.911	&	0.308 \\
Star &	UVES	&	2006-04-09	&		53834.1	&	1.493	&	-0.377	&	0.348 \\
Star &	UVES	&	2006-06-08	&		53894.0	&	1.523	&	-0.940	&	0.198 \\
Star &	UVES	&	2008-01-10	&		54475.3	&	1.810	&	-0.342	&	0.213 \\
Star &	UVES	&	2008-02-17	&		54513.3	&	1.829	&	-0.358	&	0.190 \\
Star &	UVES	&	2008-03-10	&		54535.3	&	1.839	&	-0.257	&	0.149 \\
Star &	UVES	&	2008-03-29	&		54554.3	&	1.849	&	-0.419	&	0.152 \\
Star &	UVES	&	2008-04-11	&		54567.0	&	1.855	&	-0.138	&	0.155 \\
Star &	UVES	&	2008-04-27	&		54583.0	&	1.863	&	-0.457	&	0.168 \\
Star &	UVES	&	2008-05-12	&		54599.0	&	1.871	&	-0.531	&	0.042 \\
Star &	UVES	&	2008-05-28	&		54615.0	&	1.879	&	-0.697	&	0.198 \\
Star &	UVES	&	2008-05-30	&		54616.0	&	1.879	&	-0.583	&	0.177 \\
Star &	UVES	&	2008-05-31	&		54617.1	&	1.880	&	-0.220	&	0.198 \\
Star &	UVES	&	2008-06-11	&		54629.0	&	1.886	&	-0.344	&	0.239 \\
Star &	UVES	&	2008-06-12	&		54629.0	&	1.886	&	-0.329	&	0.204 \\
Star &	UVES	&	2008-07-09	&		54656.0	&	1.899	&	-0.104	&	0.171 &	4 &	7	\\
Star &	UVES	&	2009-01-10	&		54841.4	&	1.991	&	1.306	&	0.308 &	-430 &	30	\\
Star &	UVES	&	2009-02-05	&		54867.3	&	2.004	&	0.230	&	0.030 &	77 &	 8	\\
Star &	UVES	&	2009-04-02	&		54923.2	&	2.031	&	-0.239	&	0.315 \\
Star &	UVES	&	2009-04-25	&		54946.1	&	2.043	&	-0.137	&	0.091 \\
Star &	UVES	&	2009-05-19	&		54970.0	&	2.054	&	0.0168	&	0.207 \\
Star &	UVES	&	2009-06-17	&		54999.1	&	2.069	&	-0.274	&	0.232 \\
Star &	UVES	&	2009-06-30	&		55013.0	&	2.076	&	-0.270	&	0.093 \\
Star &	UVES	&	2009-07-01	&		55014.0	&	2.076	&	-0.473	&	0.200 \\
Star &	UVES	&	2009-07-02	&		55015.0	&	2.077	&	-0.075	&	0.091 \\
Star &	UVES	&	2009-07-05	&		55018.0	&	2.078	&	-0.524	&	0.262 \\
Star &	UVES	&	2009-07-06	&		55018.0	&	2.078	&	-0.376	&	0.141 \\
Star &	UVES	&	2012-04-03	&		56020.1	&	2.573	&	-0.322	&	0.181 \\
Star &	UVES	&	2013-11-24	&		56620.3	&	2.870	&	-0.005	&	0.134 \\
Star &	UVES	&	2013-12-24	&		56650.3	&	2.885	&	-0.117	&	0.548 \\
Star &	UVES	&	2014-01-27	&		56684.2	&	2.902	&	0.129	&	0.184	& 39 &	4	\\
Star &	UVES	&	2014-02-24	&		56712.2	&	2.916	&	0.036	&	0.243	& 18 &	20	\\
Star &	UVES	&	2014-03-21	&		56737.1	&	2.928	&	0.107	&	0.171	& -21 & 23	\\
Star &	UVES	&	2014-04-16	&		56763.0	&	2.941	&	0.377	&	0.324	& 37 &	2	\\
Star &	UVES	&	2014-05-26	&		556803.1	&	2.960	&	0.925	&	0.086	&	-67 	&	56	\\
Star &	UVES	&	2014-05-27	&		556804.0	&	2.961	&	0.855	&	0.194	&	-3 & 	66	\\
Star &	UVES	&	2014-06-18	&		56826.0	&	2.972	&	1.501	&	0.230	&	-63 &	3 \\
Star &	UVES	&	2014-07-01	&		56840.0	&	2.979	&	2.547	&	0.242	&	-138 &	27	\\
Star	 &	UVES	&	2014-07-23	&		56861.0	&	2.989	&	0.476	&	0.067	&	-341 &	7	\\
Star	 &	UVES	&	2014-07-30  &		56868.0	&	2.993	&	0.147	&	0.189	&	-353 &	5 \\
Star	 &	UVES	&	2014-10-12	&	56942.4	&	3.029	&	-0.245	&	0.153	&	10	&	4	\\
Star	 &	UVES	&	2014-10-24	&	56954.4	&	3.035	&	-0.337	&	0.046	&	23	&	27	\\
Star	 &	UVES	&	2014-11-16	&	56977.3	&	3.047	&	-0.020	&	0.245	&	20	&	3	\\	
Star	 &	UVES	&	2014-12-21	&	57012.3	&	3.064	&	0.006	&	0.128		&	&	\\
Star	 &	UVES	&	2015-01-05	&	57027.4	&	3.071	&	-0.248	&	0.242	&	&	\\
Star	 &	UVES	&	2015-01-25	&	57047.3	&	3.081	&	-0.164	&	0.241	&	&		\\
		\hline
Star & CHIRON & 2013-05-16 & 56428.1 & 2.775 & 0.503 & 0.061 & 422 & 18 \\ 
Star & CHIRON & 2013-11-11 & 56607.3 & 2.864 & 0.735 & 0.026 & 32 & 1 \\ 
Star & CHIRON & 2014-01-20 & 56677.4 & 2.898 & 0.896 & 0.025 & 80 & 1 \\ 
Star & CHIRON & 2014-01-21 & 56678.3 & 2.899 & 0.594 & 0.033 & 59 & 14 \\ 
Star & CHIRON & 2014-02-10 & 56698.3 & 2.909 & 0.852 & 0.003 & 74 & 3 \\ 
Star & CHIRON & 2014-03-08 & 56724.1 & 2.921 & 1.184 & 0.042 & -94 & 8 \\ 
Star & CHIRON & 2014-04-03 & 56750.0 & 2.934 & 1.436 & 0.027 & 30 & 3 \\ 
Star & CHIRON & 2014-04-27 & 56775.0 & 2.947 & 1.417 & 0.138 & -128 & 2 \\ 
Star & CHIRON & 2014-04-30 & 56777.0 & 2.948 & 1.266 & 0.124 & -145 & 13 \\ 
Star & CHIRON & 2014-05-09 & 56786.0 & 2.952 & 1.297 & 0.141 & -64 & 18 \\ 
Star & CHIRON & 2014-05-10 & 56787.0 & 2.953 & 1.447 & 0.141 & -73 & 12 \\ 
Star & CHIRON & 2014-05-17 & 56795.0 & 2.956 & 1.330 & 0.147 & -179 & 3 \\ 
Star & CHIRON & 2014-05-25 & 56802.0 & 2.960 & 1.726 & 0.144 & -88 & 30 \\ 
Star & CHIRON & 2014-05-29 & 56806.0 & 2.962 & 1.295 & 0.156 & -82 & 35 \\ 
Star & CHIRON & 2014-05-30 & 56807.1 & 2.962 & 1.317 & 0.134 & -8 & 1 \\ 
Star & CHIRON & 2014-06-01 & 56809.0 & 2.963 & 1.765 & 0.150 & -91 & 24 \\ 
Star & CHIRON & 2014-06-03 & 56811.0 & 2.964 & 1.622 & 0.159 & -135 & 2 \\ 
Star & CHIRON & 2014-06-11 & 56818.0 & 2.968 & 1.650 & 0.144 & -209 & 33 \\ 
Star & CHIRON & 2014-06-18 & 56825.0 & 2.971 & 2.340 & 0.215 & -95 & 16 \\ 
Star & CHIRON & 2014-06-21 & 56829.0 & 2.973 & 2.311 & 0.258 & -26 & 15 \\ 
Star & CHIRON & 2014-06-26 & 56833.0 & 2.975 & 1.971 & 0.234 & -64 & 12 \\ 
Star & CHIRON & 2014-06-29 & 56837.0 & 2.977 & 2.306 & 0.231 & -203 & 25 \\ 
Star & CHIRON & 2014-07-01 & 56839.0 & 2.978 & 2.577 & 0.254 & -230 & 7 \\ 
Star & CHIRON & 2014-07-02 & 56840.0 & 2.979 & 3.454 & 0.195 & -161 & 30 \\ 
Star & CHIRON & 2014-07-09 & 56847.0 & 2.982 & 2.422 & 0.174 & -209 & 3 \\ 
Star & CHIRON & 2014-07-18 & 56856.0 & 2.987 & 4.000 & 0.333 & -303 & 20 \\ 
Star & CHIRON & 2014-07-19 & 56857.0 & 2.987 & 4.282 & 0.335 & -267 & 27 \\ 
Star & CHIRON & 2014-07-20 & 56858.0 & 2.988 & 3.831 & 0.247 & -321 & 12 \\ 
Star & CHIRON & 2014-07-21 & 56859.0 & 2.988 & 3.538 & 0.257 & -286 & 14 \\ 
Star & CHIRON & 2014-07-25 & 56863.0 & 2.990 & 3.478 & 0.248 & -366 & 15 \\ 
Star & CHIRON & 2014-07-26 & 56864.0 & 2.991 & 3.260 & 0.267 & -381 & 4 \\ 
Star & CHIRON & 2014-07-28 & 56866.0 & 2.992 & 1.637 & 0.157 & -390 & 20 \\ 
Star & CHIRON & 2014-07-29 & 56867.0 & 2.992 & 1.795 & 0.136 & -431 & 2 \\ 
Star & CHIRON & 2014-08-01 & 56870.0	&	2.994 & 0.670 & 0.035 &	-124 & 12	\\
Star & CHIRON & 2014-08-02 & 56871.0	&	2.994 & 0.851 & 0.025 &	-25 &	120	\\
Star & CHIRON & 2014-08-14 & 56883.0	&	3.000 & 1.337 & 0.178 &	-74 & 66	\\
Star & CHIRON & 2014-08-16 & 56885.0	&	3.001 & 1.259 & 0.155 &	-102 &	16	\\
Star & CHIRON & 2014-08-17 & 56886.0	&	3.001	& 1.152 & 0.159 &	-115 & 2\\
Star & CHIRON & 2014-08-18 & 56887.0	&	3.002	&	0.903	&	0.124 &	-41	&	6	\\
Star & CHIRON & 2014-10-21	&	56951.4	&	3.034	&	0.863	&	0.075	&	58	& 20 	\\
Star & CHIRON & 2014-10-22	&	56952.4	&	3.034	&	0.714	&	0.105	&	78	&	7	\\
Star & CHIRON & 2014-10-23	&	56953.4	&	3.035	&	0.647	&	0.123	&	74	&	1	\\
Star & CHIRON & 2014-10-24	&	56954.3	&	3.035	&	0.740	&	0.040	&	75	&	1	\\
Star & CHIRON & 2014-10-25	&	56955.4	&	3.036	&	0.817	&	0.084	&	69	&	4	\\
Star & CHIRON & 2014-10-26	&	56956.4	&	3.036	&	0.659	&	0.085	&	56	&	3	\\
Star & CHIRON & 2014-10-27	&	56957.4	&	3.037	&	0.528	&	0.112	&	57	&	11	\\				
\end{longtable}

\begin{longtable}{llcccccrr}
\caption{\label{table:heIIfos4} \ion{He}{II} equivalent width and velocity at FOS4. Velocity is the peak velocity and is corrected for the moving-mirror effect.}\\
\hline\hline
Position &	Instrument	& Date & MJD  & Phase  & $-$EW & $\sigma$EW   & Vel & $\sigma$Vel  \\ 
 &	& & (days)  &  & (\AA)& (\AA) & (km~s$^{-1}$)& (km~s$^{-1}$) \\ 
\hline
\endfirsthead
\caption{continued.}\\
\hline\hline
Location &	Instrument	& Date & MJD  & Phase  & $-$EW & $\sigma$EW   & Vel & $\sigma$Vel    \\ 
& & & (days)  &  & (\AA)& (\AA) & (km~s$^{-1}$)& (km~s$^{-1}$) \\ 
\hline
\endhead
\hline
\endfoot
FOS4	&	UVES	&	2002-12-26	&	52634.4	&	0.900	&	-0.107	&	0.139 \\
FOS4	&	UVES	&	2002-12-31	&	52639.3	&	0.902	&	0.066	&	0.116 \\
FOS4	&	UVES	&	2003-01-03	&	52642.3	&	0.904	&	-0.053	&	0.115 \\
FOS4	&	UVES	&	2003-01-23	&	52662.4	&	0.914	&	0.008	&	0.060 \\
FOS4	&	UVES	&	2003-02-04	&	52674.4	&	0.920	&	0.005	&	0.090 \\
FOS4	&	UVES	&	2003-02-14	&	52684.1	&	0.924	&	0.151	&	0.100 \\
FOS4	&	UVES	&	2003-02-25	&	52695.3	&	0.930	&	0.146	&	0.099 \\
FOS4	&	UVES	&	2003-03-12	&	52710.0	&	0.937	&	0.387	&	0.103 \\
FOS4	&	UVES	&	2003-04-30	&	52759.1	&	0.962	&	0.231	&	0.143 &	48	& 17	\\
FOS4	&	UVES	&	2003-05-05	&	52765.0	&	0.964	&	0.348	&	0.098 &	-85 &	9	\\
FOS4	&	UVES	&	2003-05-12	&	52771.2	&	0.967	&	0.378	&	0.115 &	-126 &	1	\\
FOS4	&	UVES	&	2003-05-29	&	52788.0	&	0.976	&	0.515	&	0.162 &	-55 &	24	\\
FOS4	&	UVES	&	2003-06-03	&	52794.0	&	0.979	&	0.627	&	0.173 &	-58 &	2	\\
FOS4	&	UVES	&	2003-06-08	&	52798.0	&	0.981	&	0.720	&	0.158 &	-50 &	9	\\
FOS4	&	UVES	&	2003-06-13	&	52803.0	&	0.983	&	0.820	&	0.131 &	-74 &	6	\\
FOS4	&	UVES	&	2003-06-17	&	52808.0	&	0.986	&	0.958	&	0.207 &	-183 &	68	\\
FOS4	&	UVES	&	2003-06-22	&	52813.0	&	0.988	&	0.971	&	0.178 &	-100 &	16	\\
FOS4	&	UVES	&	2003-06-30	&	52821.0	&	0.992	&	1.068	&	0.318 &	-107 &	17	\\
FOS4	&	UVES	&	2003-07-05	&	52825.5	&	0.994	&	0.736	&	0.568 &	-232 &	29	\\
FOS4	&	UVES	&	2003-07-09	&	52830.0	&	0.997	&	1.278	&	0.423 &	-370 &	8	\\
FOS4	&	UVES	&	2003-07-21	&	52841.0	&	1.002	&	0.363	&	0.564 	\\
FOS4	&	UVES	&	2003-07-26	&	52847.0	&	1.005	&	0.493	&	0.188 \\
FOS4	&	UVES	&	2003-07-27	&	52848.0	&	1.005	&	0.770	&	1.371	\\
FOS4	&	UVES	&	2003-08-01	&	52852.0	&	1.007	&	0.463	&	0.955 \\
FOS4	&	UVES	&	2003-11-25	&	52968.3	&	1.065	&	0.423	&	0.519 \\
FOS4	&	UVES	&	2003-12-17	&	52990.3	&	1.076	&	0.417	&	0.491 &	-53 &	28	\\
FOS4	&	UVES	&	2004-01-02	&	53006.3	&	1.084	&	0.334	&	0.421 \\
FOS4	&	UVES	&	2004-01-25	&	53029.3	&	1.095	&	0.249	&	0.326 \\
FOS4	&	UVES	&	2004-02-20	&	53055.2	&	1.108	&	0.255	&	0.339 \\
FOS4	&	UVES	&	2004-03-11	&	53075.1	&	1.118	&	0.240	&	0.255 \\
FOS4	&	UVES	&	2004-12-10	&	53349.3	&	1.253	&	0.108	&	0.197 \\
FOS4	&	UVES	&	2005-01-19	&	53389.2	&	1.273	&	0.154	&	0.277 \\
FOS4	&	UVES	&	2005-03-02	&	53431.3	&	1.294	&	-0.089	&	0.252 \\
FOS4	&	UVES	&	2006-05-11	&	53866.0	&	1.509	&	-0.107	&	0.388 \\
FOS4	&	UVES	&	2006-06-26	&	53912.1	&	1.531	&	-0.174	&	0.395 \\
FOS4	&	UVES	&	2008-02-17	&	54513.3	&	1.829	&	0.096	&	0.037 \\
FOS4	&	UVES	&	2008-03-29	&	54554.3	&	1.849	&	0.064	&	0.019 \\
FOS4	&	UVES	&	2008-04-11	&	54567.0	&	1.855	&	0.105	&	0.043 \\
FOS4	&	UVES	&	2008-04-27	&	54583.0	&	1.863	&	0.065	&	0.009 \\
FOS4	&	UVES	&	2008-05-12	&	54599.0	&	1.871	&	0.015	&	0.040 \\
FOS4	&	UVES	&	2008-05-30	&	54616.1	&	1.879	&	0.011	&	0.017 \\
FOS4	&	UVES	&	2008-05-31	&	54617.1	&	1.880	&	0.116	&	0.016 \\
FOS4	&	UVES	&	2008-06-11	&	54629.0	&	1.886	&	0.118	&	0.015 &	-35	& 9	\\
FOS4	&	UVES	&	2008-07-09	&	54656.1	&	1.899	&	0.198	&	0.026 &	-53 &	45	\\
FOS4	&	UVES	&	2008-07-10	&	54657.1	&	1.900	&	0.109	&	0.027 &	-14 &	31	\\
FOS4	&	UVES	&	2009-01-25	&	54856.2	&	1.998	&	1.773	&	0.339 &	-350 &	4	\\
FOS4	&	UVES	&	2009-02-20	&	54882.2	&	2.011	&	1.127	&	0.655 &	-33 &	9	\\
FOS4	&	UVES	&	2009-04-02	&	54923.3	&	2.031	&	0.295	&	0.269 &	56 &	 32	\\
FOS4	&	UVES	&	2009-04-25	&	54946.1	&	2.043	&	0.294	&	0.263 &	9 &	2	\\
FOS4	&	UVES	&	2009-06-17	&	54999.1	&	2.069	&	0.354	&	0.188 &	-74 &	13	\\
FOS4	&	UVES	&	2009-06-30	&	55013.0	&	2.076	&	0.055	&	0.115 \\
FOS4	&	UVES	&	2009-07-01	&	55013.5	&	2.076	&	0.050	&	0.099 \\
FOS4	&	UVES	&	2009-07-02	&	55014.7	&	2.076	&	0.164	&	0.047 \\
FOS4	&	UVES	&	2009-07-05	&	55018.0	&	2.078	&	0.128	&	0.083 \\
FOS4	&	UVES	&	2009-07-06	&	55018.0	&	2.078	&	0.169	&	0.094 \\
FOS4	&	UVES	&	2012-04-04	&	56021.2	&	2.574	&	0.059	&	0.073 \\
FOS4	&	UVES	&	2013-11-24	&	56620.3	&	2.870	&	0.298	&	0.011 \\
FOS4	&	UVES	&	2013-12-24	&	56650.3	&	2.885	&	0.081	&	0.140 \\
FOS4	&	UVES	&	2014-01-27	&	56684.2	&	2.902	&	0.296	&	0.011 &	-26 &	4	\\
FOS4	&	UVES	&	2014-02-24	&	56712.2	&	2.916	&	0.118	&	0.052 \\
FOS4	&	UVES	&	2014-03-21	&	56737.2	&	2.928	&	0.017	&	0.029 &	-41 &	18	\\
FOS4	&	UVES	&	2014-04-16	&	56763.1	&	2.941	&	0.297	&	0.058 &	-107 &	7	\\
FOS4	&	UVES	&	2014-05-26	&	56804.0	&	2.961	&	0.535	&	0.003 &	-150 &	17	\\
FOS4	&	UVES	&	2014-06-18	&	56826.1	&	2.972	&	0.663	&	0.065 &	-91 &	6	\\	
FOS4	&	UVES	&	2014-07-01	&	56840.0	&	2.979	&	1.085	&	0.109 &	-32 &	47	\\ 
FOS4&	UVES	&	2014-07-27	&	56865.0	&	2.991	&	1.778	&	0.080 &	-148 &	10	\\
FOS4&	UVES	&	2014-07-30	&	56868.0	&	2.993	&	1.627	&	0.135	&	-86 &	23	\\
FOS4&	UVES	&	2014-10-12	&	56942.4	&	3.029	&	0.607	&	0.239	&	-30	&	50	\\
FOS4&	UVES	&	2014-10-24	&	56954.4	&	3.035	&	0.469	&	0.217	&	-17	&	72	\\
FOS4&	UVES	&	2014-11-16	&	56977.3	&	3.047	&	0.547	&	0.225	&	-127 &	11 \\
FOS4&	UVES	&	2014-12-08	&	56999.3	&	3.058	&	0.382	&	0.137	&	-17 &	59 \\
FOS4&	UVES	&	2014-12-21	&	57012.3	&	3.064	&	0.205	&	0.193	&	&	\\
FOS4&	UVES	&	2015-01-01	&	57027.3	&	3.071	&	0.283	&	0.172	&	&	\\
FOS4&	UVES	&	2015-01-21	&	57043.4	&	3.079	&	0.112	&	0.098	&	&	\\
FOS4&	UVES	&	2015-01-25	&	57047.3	&	3.081	&	0.165	&	0.028	&	&	\\										

\end{longtable}

\end{document}